\documentstyle[preprint,eqsecnum,prd,aps,epsf]{revtex}
\tightenlines
\begin{document}
\draft 


\newcommand{\fix}[1]{{\bf <<< #1 !!! }}
\newcommand\mutrue{\mu_t} 
\newcommand\mubest{\mu_{\rm best}}
\newcommand\noise{\varepsilon}
\newcommand\Rnew{\widetilde R}
\newcommand\Cnew{\widetilde c}
\newcommand\erf{{\mathrm{erf}}}

\draft
\title{Application of Conditioning to the Gaussian-with-Boundary
Problem in the Unified Approach to Confidence Intervals}
\author{Robert D. Cousins\cite{bylineRC}}
\address{Department of Physics and Astronomy, University of California,
Los Angeles, CA 90095}
\date{January 14, 2000}
\maketitle
\begin{abstract}
Roe and Woodroofe (RW) have suggested that certain conditional
probabilities be incorporated into the ``unified approach'' for
constructing confidence intervals, previously described by Feldman and
Cousins (FC).  RW illustrated this conditioning technique using one of
the two prototype problems in the FC paper, that of Poisson processes
with background.  The main effect was on the upper curve in the
confidence belt.  In this paper, we attempt to apply this style of
conditioning to the other prototype problem, that of Gaussian errors
with a bounded physical region.  We find that the lower curve on the
confidence belt is also moved significantly, in an undesirable manner.
\end{abstract}


\narrowtext

\section{Introduction}
Roe and Woodroofe \cite{RW} have made an interesting suggestion for
modifying the ``unified approach'' to classical confidence intervals
which Feldman and I advocated in Ref.~\cite{FC}.  They invoke the use
of ``conditioning'', namely replacing frequentist coverage
probabilities with conditional probabilities, still calculated in a
frequentist manner, but conditioned on knowledge gained from the
result of the particular experiment at hand.

Roe and Woodroofe (RW) illustrate their suggestion using one of the
two prototype problems, that of Poisson processes with background.
Suppose, for example that an experiment observes 3 events (signal plus
background).  Then the experimenters know that, in that particular
experiment, there were 3 or fewer background events.  RW therefore
calculate the frequentist coverage using an ensemble of experiments
with 3 or fewer background events, rather than the larger unrestricted
ensemble which we used.  Thus, the RW ensemble changes from experiment
to experiment.  Conditioning on an equality has a long history in
classical statistics. (Ref.~\cite{RW} contains key references.)
However, conditioning on an inequality, as RW do when the number of
events is greater than zero, is perhaps less well founded, and it is
interesting to explore the consequences.

In this paper, we attempt to apply RW-like conditioning to the other
prototype problem, that of Gaussian errors with a bounded physical
region.  The result is similar to the Poisson problem analyzed by RW,
but difficulties which were apparently masked by the discrete nature
of the Poisson problem now arise.  In particular, the lower endpoints
of confidence intervals are moved significantly in an undesirable
direction.

\section{The Unified Approach to the Gaussian-with-Boundary
Problem}

As in Ref.~\cite{FC}, we consider an observable $x$ which is the
measured value of parameter $\mu$ in an experiment with a Gaussian
resolution function with known fixed rms deviation $\sigma$, set here
to unity.  I.e.,
\begin{equation}
\label{eqn-gauss}
P(x|\mu)  = {1\over\sqrt{2\pi}}\exp(-(x-\mu)^2/2).
\end{equation}
We consider the interesting case where only non-negative values for
$\mu$ are physically allowed (for example, if $\mu$ is a
mass). 

The confidence-belt construction  in Ref.~\cite{FC} proceeded as
follows.  For a particular $x$, we let $\mubest$ be the
physically allowed value of $\mu$ for which $P(x|\mu)$ is
maximum.  Then $\mubest = \max(0,x)$, and
\begin{equation}
\label{eqn-pmubest}
P(x|\mubest) = \left\{ \begin{array}{ll}
                       1/\sqrt{2\pi}, & \mbox{$x\ge0$}\\
                       \exp(-x^2/2)/\sqrt{2\pi}, & \mbox{$x<0$.}
                       \end{array}
               \right. 
\end{equation}
We then compute the likelihood ratio $R$,
\begin{equation}
\label{eqn-R-gauss}
R(x) = {P(x|\mu) \over P(x|\mubest) }
         = \left\{ \begin{array}{ll}
                   \exp(-(x-\mu)^2/2), & \mbox{$x\ge0$}\\
                   \exp(x\mu - \mu^2/2), & \mbox{$x<0$.}
                   \end{array}
           \right.
\end{equation}
During our Neyman construction of confidence intervals, $R$ determines
the order in which values of $x$ are added to the acceptance region at
a particular value of $\mu$.  In practice, this means that for a
given value of $\mu$, one finds the interval $[x_1,x_2]$ such that
$R(x_1)=R(x_2)$ and
\begin{equation}
\label{eqn-accept}
\int_{x_1}^{x_2} P(x|\mu) dx = \alpha,
\end{equation}
where $\alpha$ is the confidence level (C.L.).
We solve for $x_1$ and $x_2$ numerically to the desired precision, for
each $\mu$ in a fine grid.  With the acceptance regions
all constructed, we then read off the confidence intervals
$[\mu_1,\mu_2]$ as in Ref.~\cite{FC}.

\section{Invoking Conditioning in the  Gaussian-with-Boundary
Problem}

In order to formulate the conditioning, we find it helpful to
think of the measured value $x$ as being the sum of two
parts, the true mean $\mutrue$ and the random ``noise'' which we
call $\noise$:
\begin{equation}
\label{eqn-epsilon}
x = \mutrue + \noise.
\end{equation}
We are considering the case where it is known on physical grounds that
$\mutrue \ge 0$.  
Thus, if an experimenter obtains the value $x_0$ in an particular
experiment, then he or she knows that, {\em in that particular experiment},
\begin{equation}
\label{eqn-epsilon-ineq}
\noise \le x_0.
\end{equation}
For example, if the experimenter measures $\mu$ and obtains $x_0 =
-2$, then the experimenter knows that $\noise \le -2$ in that
particular experiment.  This information is analogous to the
information in the Poisson problem above in which one knows that in
the particular experiment, the number of background events is 3 or
fewer.  We thus use it the manner analogous to that of RW: our
particular experimenter will consider the ensemble of experiments with
$\noise \le x_0$ when constructing the confidence belt relevant to his
or her experiment.  

\begin{figure}
\begin{center}
  \begin{tabular}{ccc}
   \leavevmode \epsfxsize=5cm \epsfbox{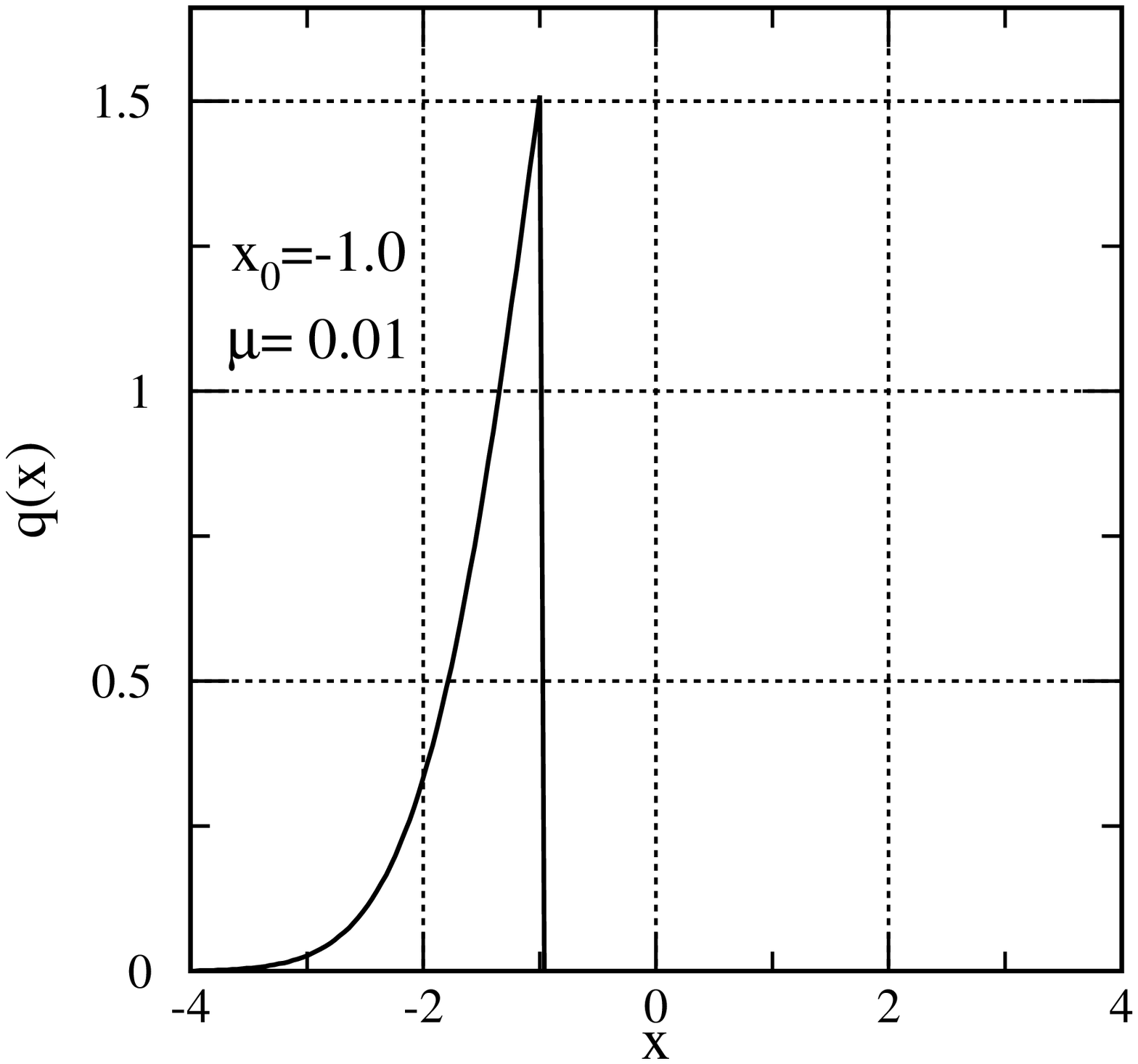} &
   \leavevmode \epsfxsize=5cm \epsfbox{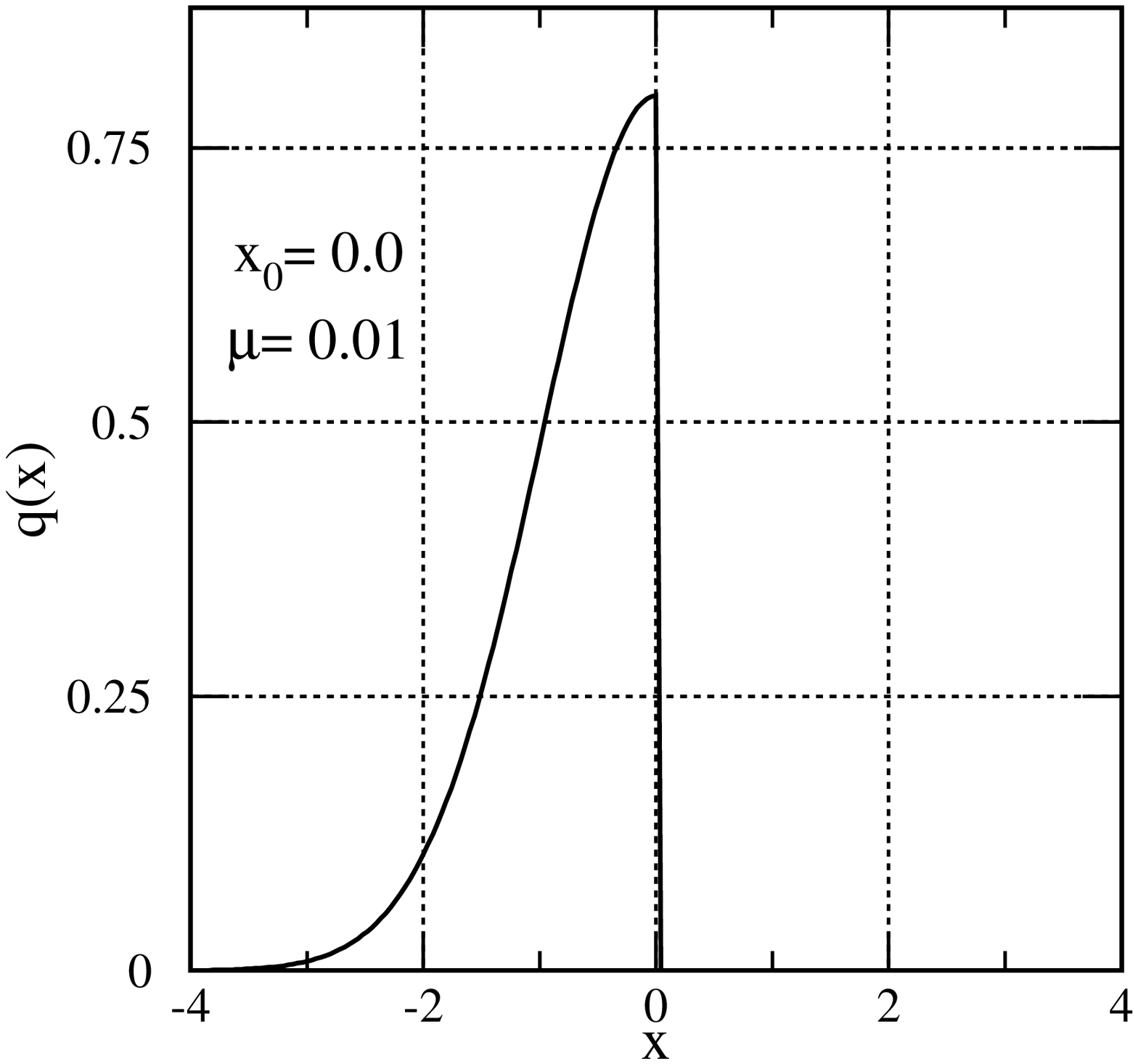} &
   \leavevmode \epsfxsize=5cm \epsfbox{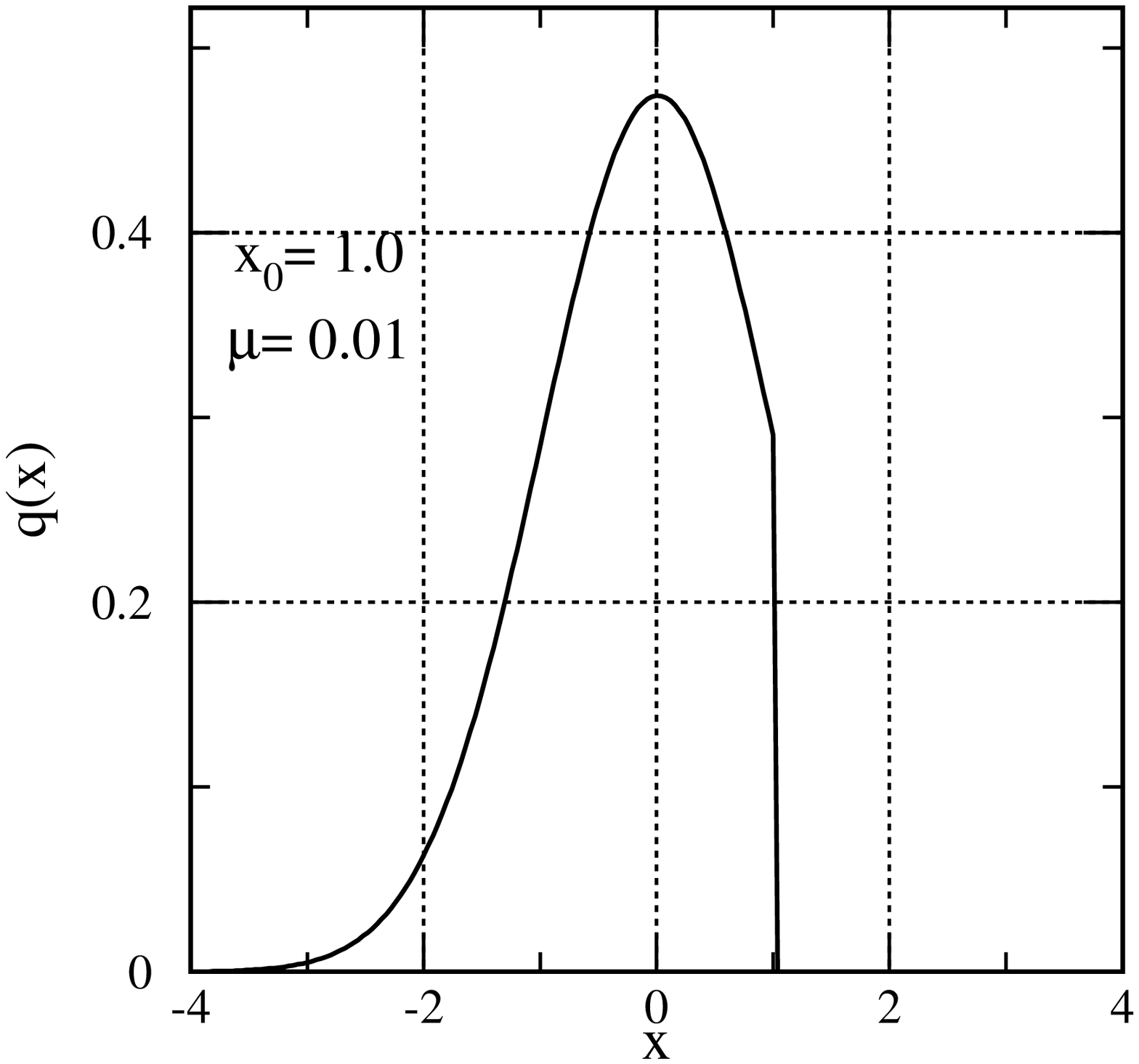} \cr
   \leavevmode \epsfxsize=5cm \epsfbox{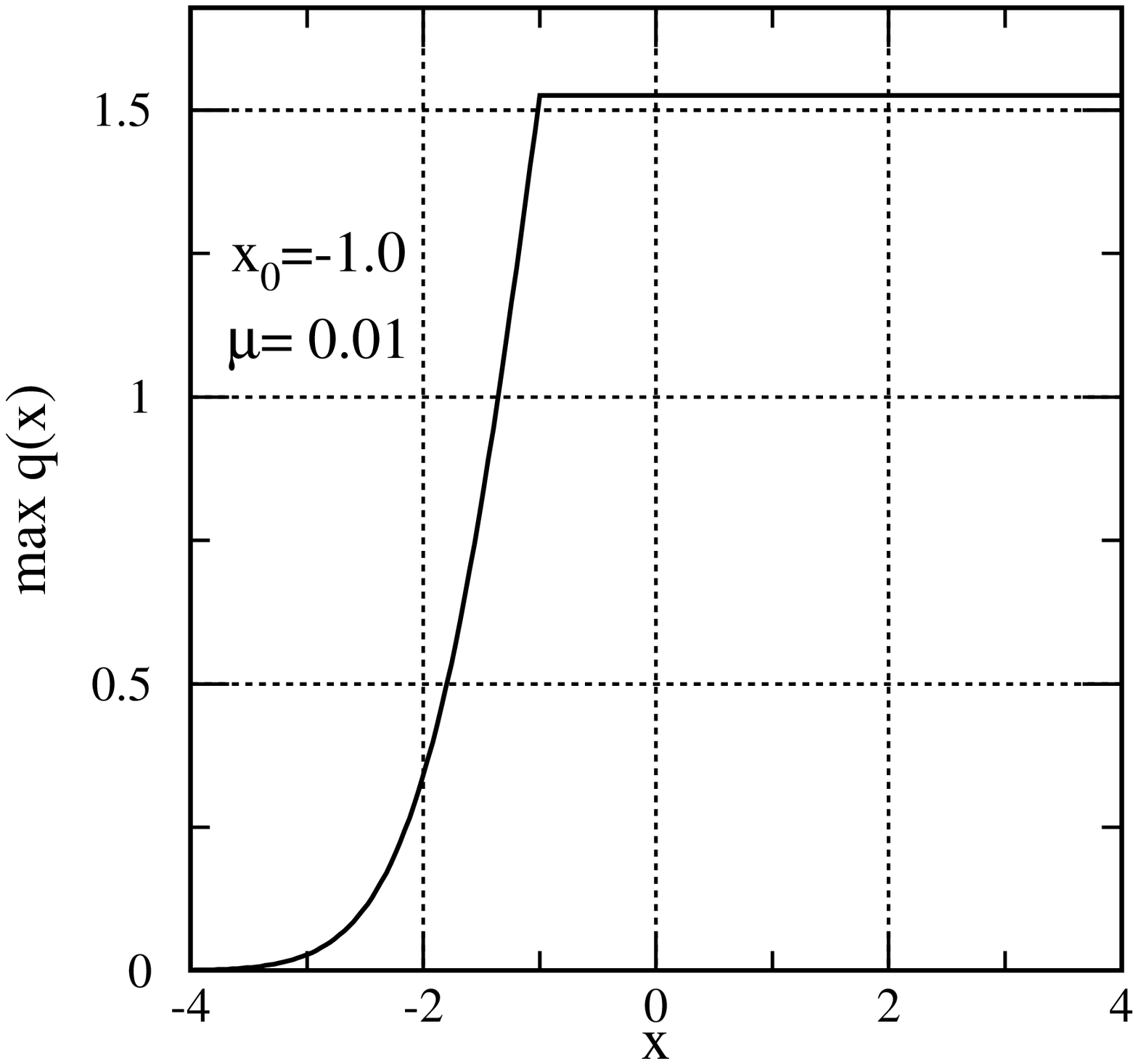} &
   \leavevmode \epsfxsize=5cm \epsfbox{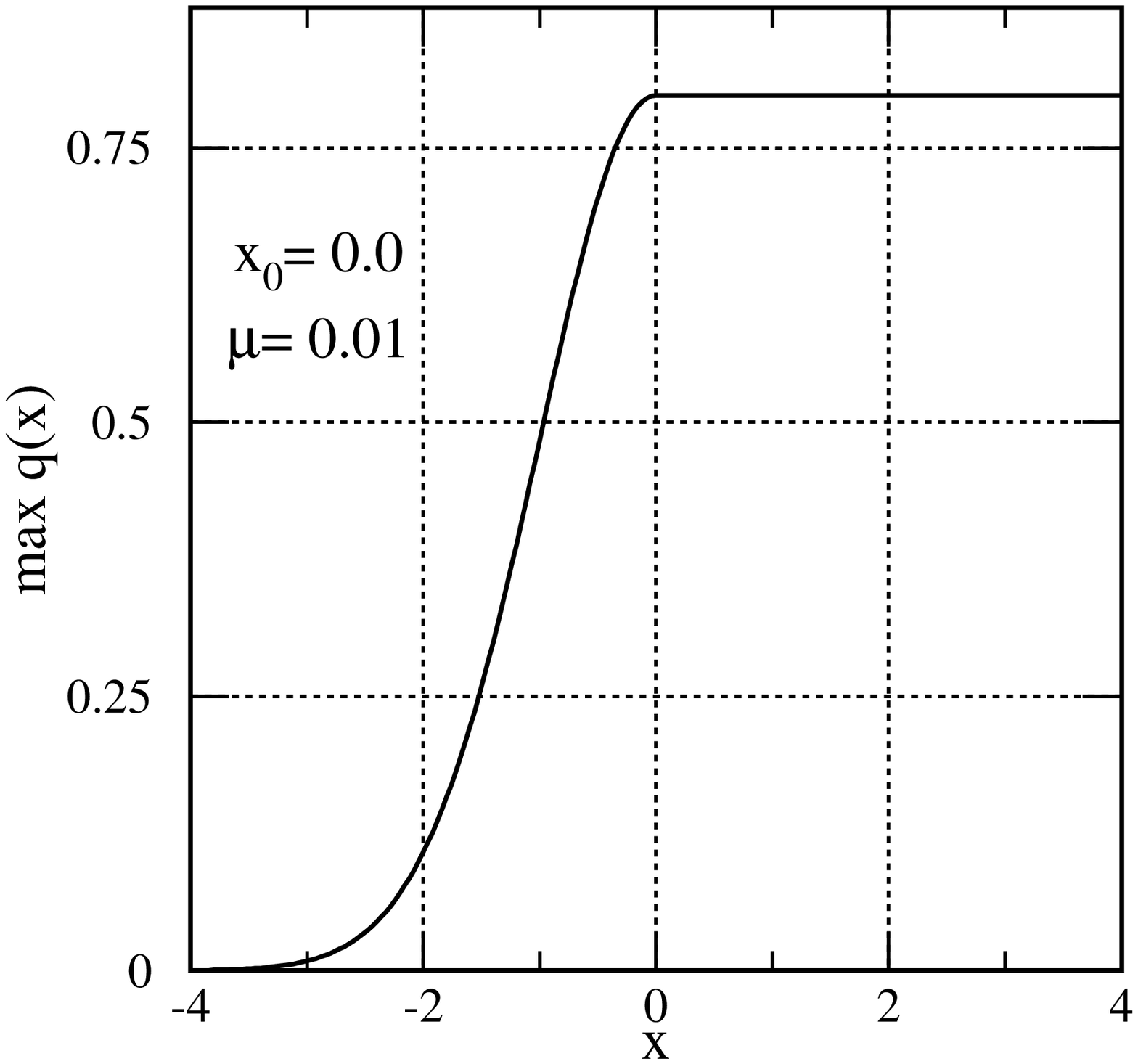} &
   \leavevmode \epsfxsize=5cm \epsfbox{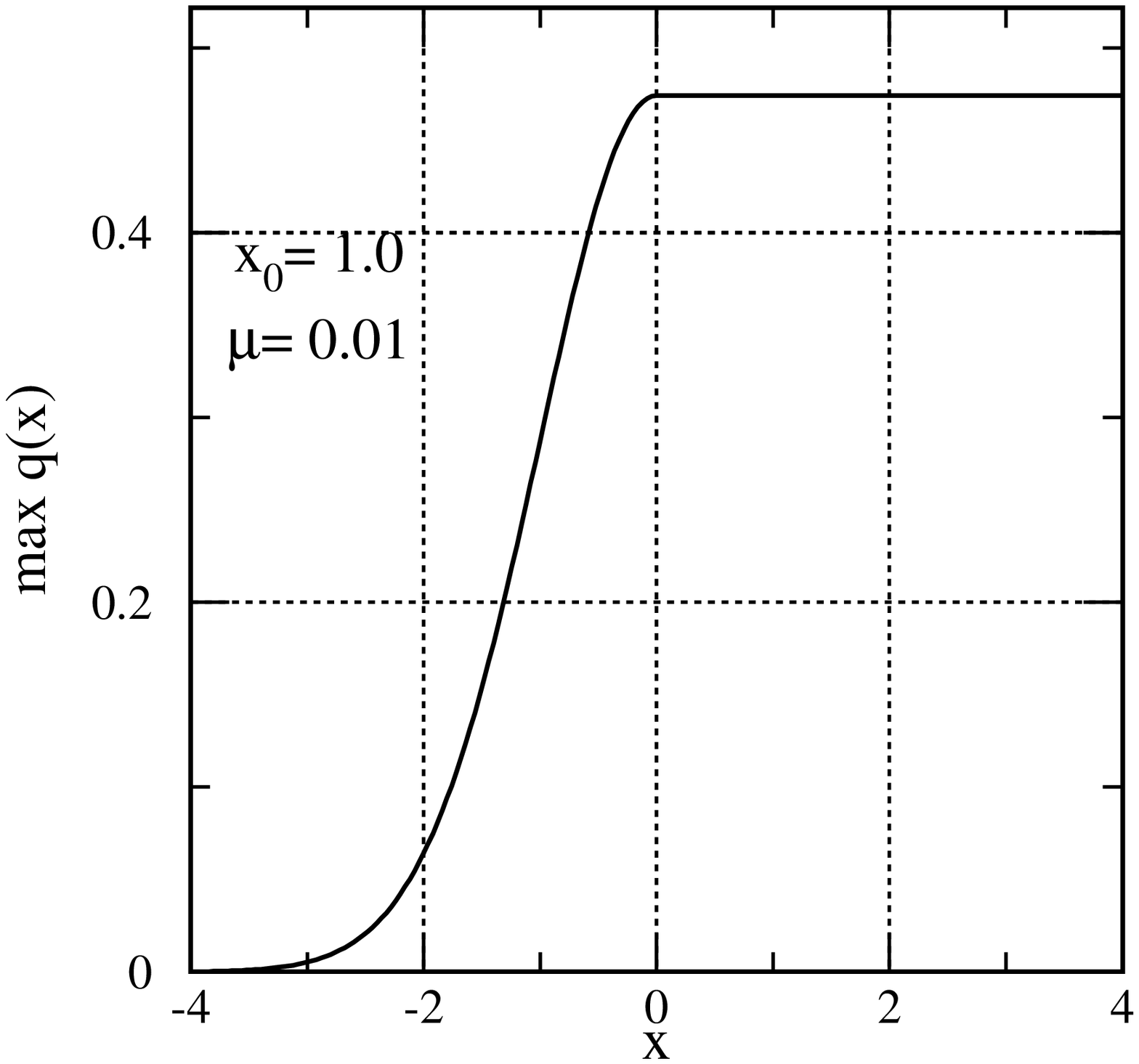} \cr
   \leavevmode \epsfxsize=5cm \epsfbox{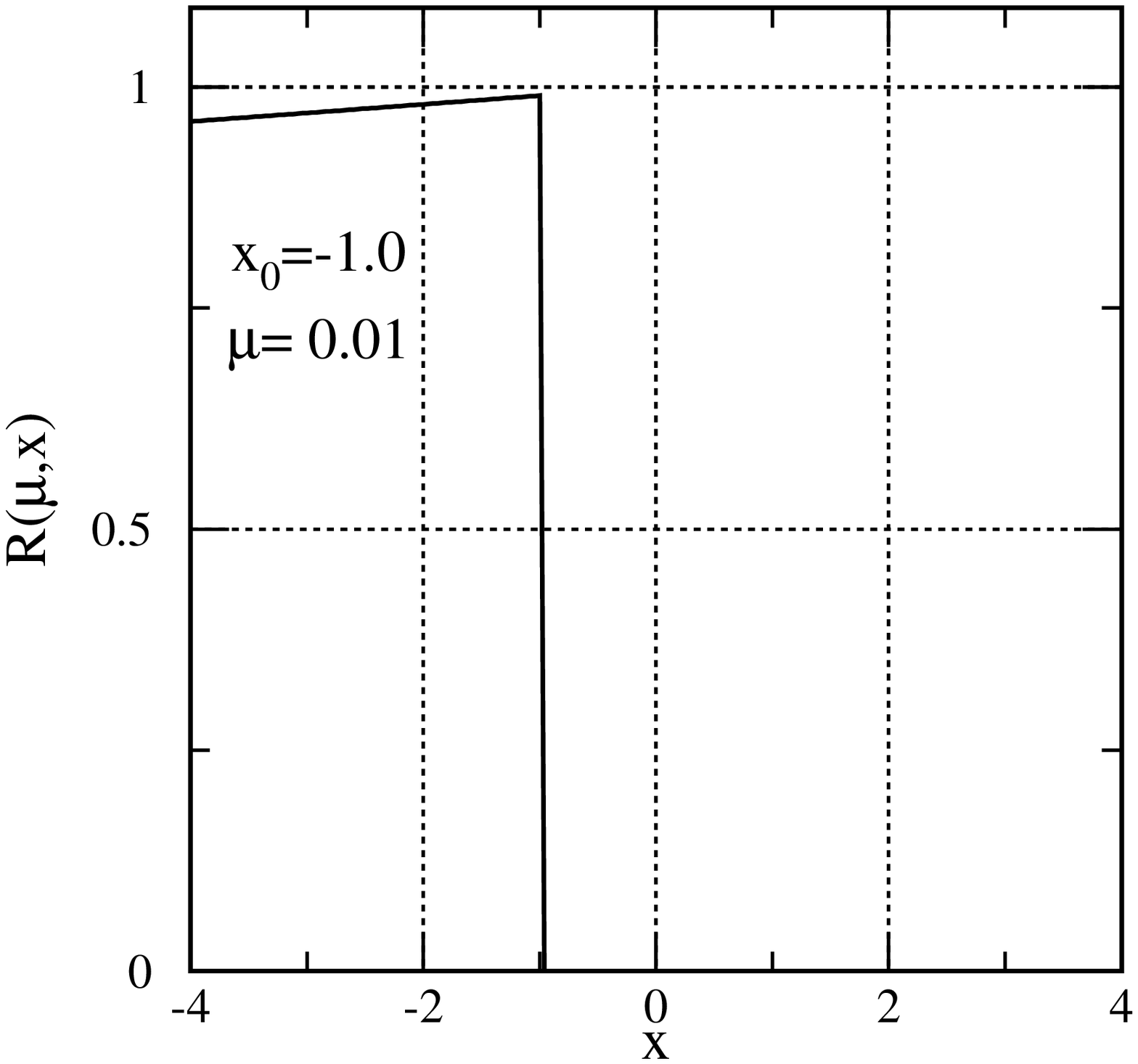} &
   \leavevmode \epsfxsize=5cm \epsfbox{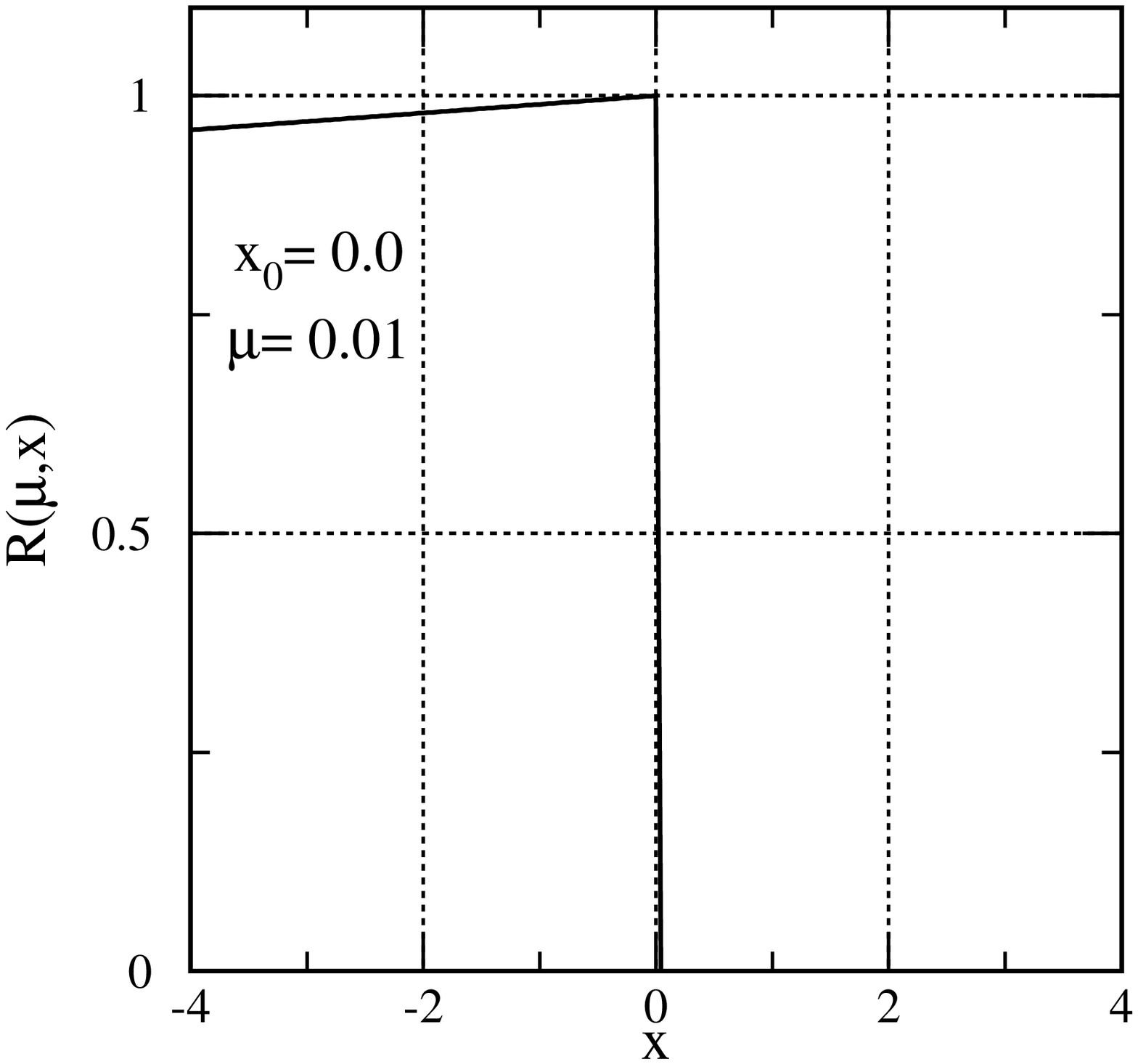} &
   \leavevmode \epsfxsize=5cm \epsfbox{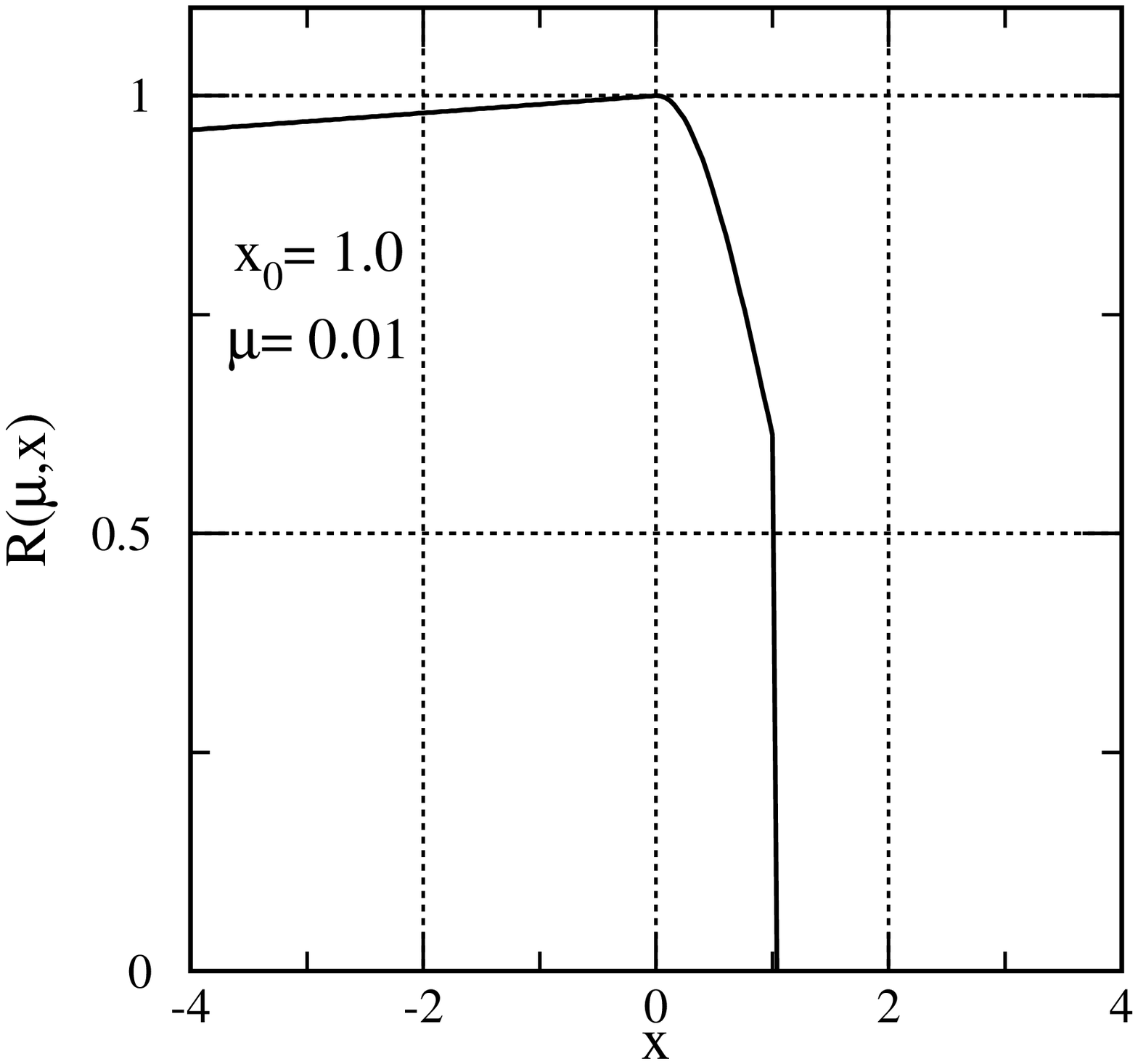} \cr
  \end{tabular}
\end{center}
\caption{Graphs of $ q^{x_0}_\mu(x)$ (top row), 
$\max_{\mu^\prime}\,q^{x_0}_{\mu^\prime}(x)$ (middle row), 
and $\Rnew^{x_0}(\mu,x)$ (bottom row), for $\mu=0.01$.
The columns are for $x_0$ = $-$1, 0, and 1.  Each graph in the bottom
row is the quotient of the two graphs above it.}
\label{fig-mu0.01}
\end{figure}

\begin{figure}
\begin{center}
  \begin{tabular}{ccc}
   \leavevmode \epsfxsize=5cm \epsfbox{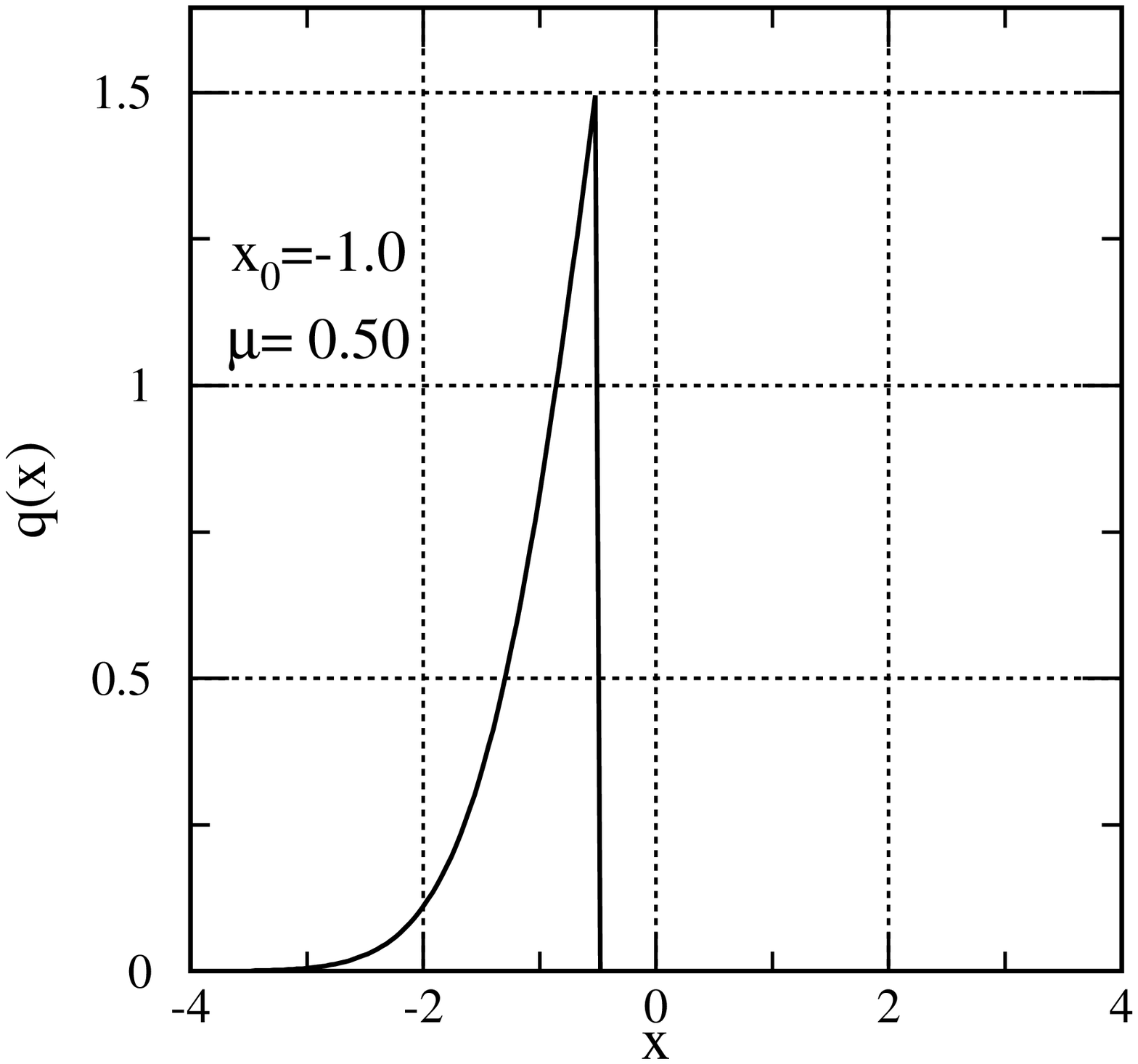} &
   \leavevmode \epsfxsize=5cm \epsfbox{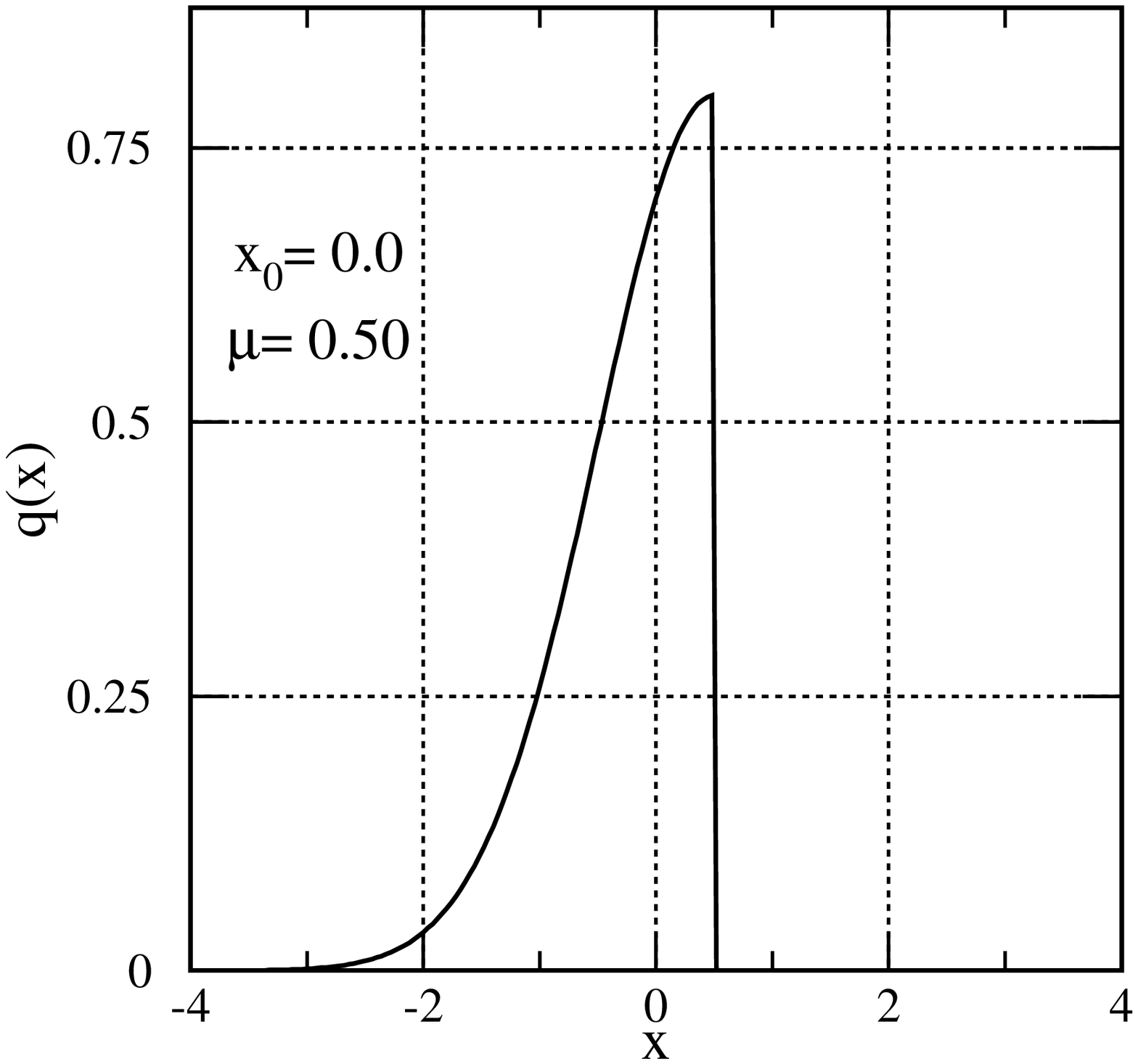} &
   \leavevmode \epsfxsize=5cm \epsfbox{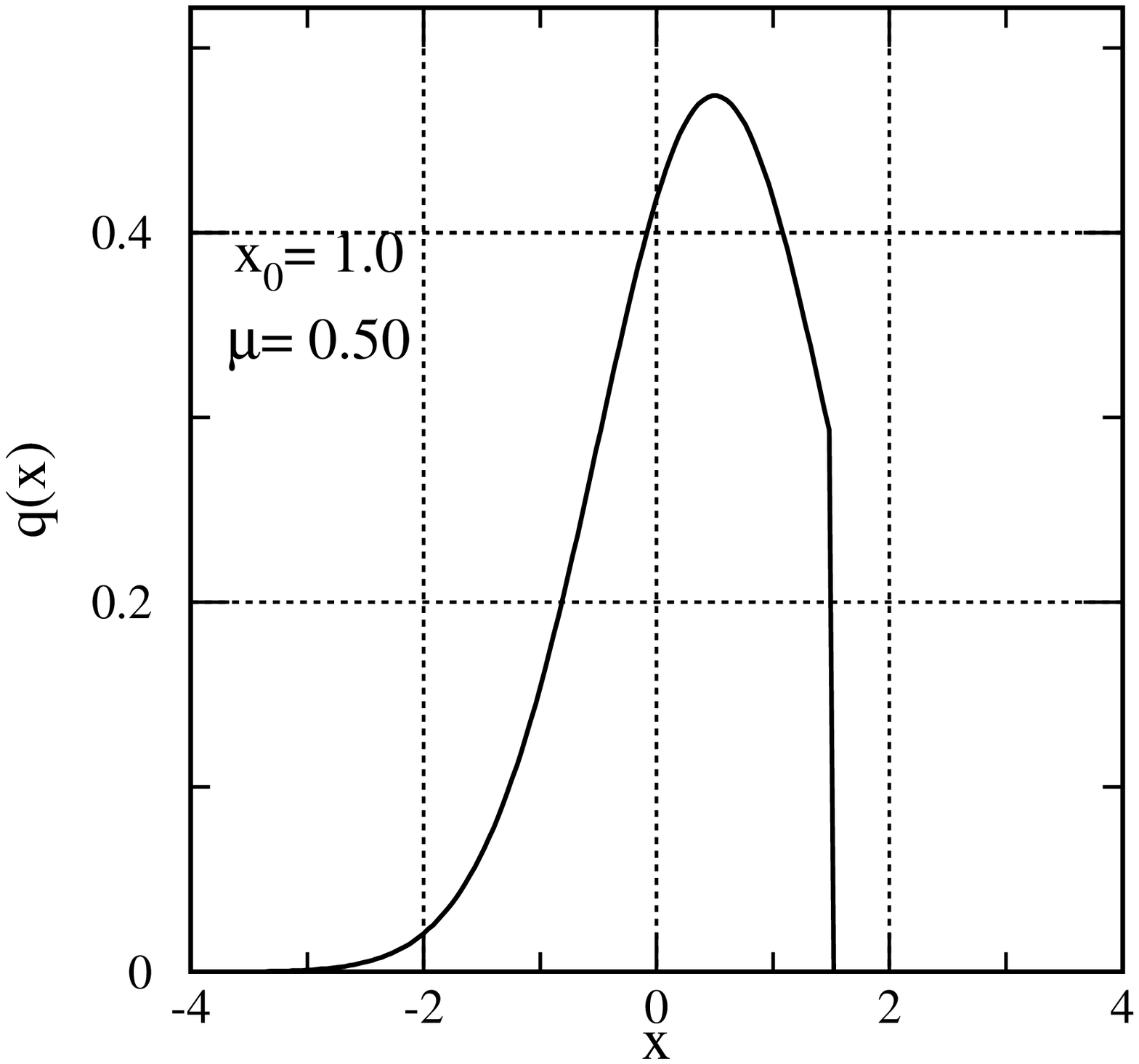} \cr
   \leavevmode \epsfxsize=5cm \epsfbox{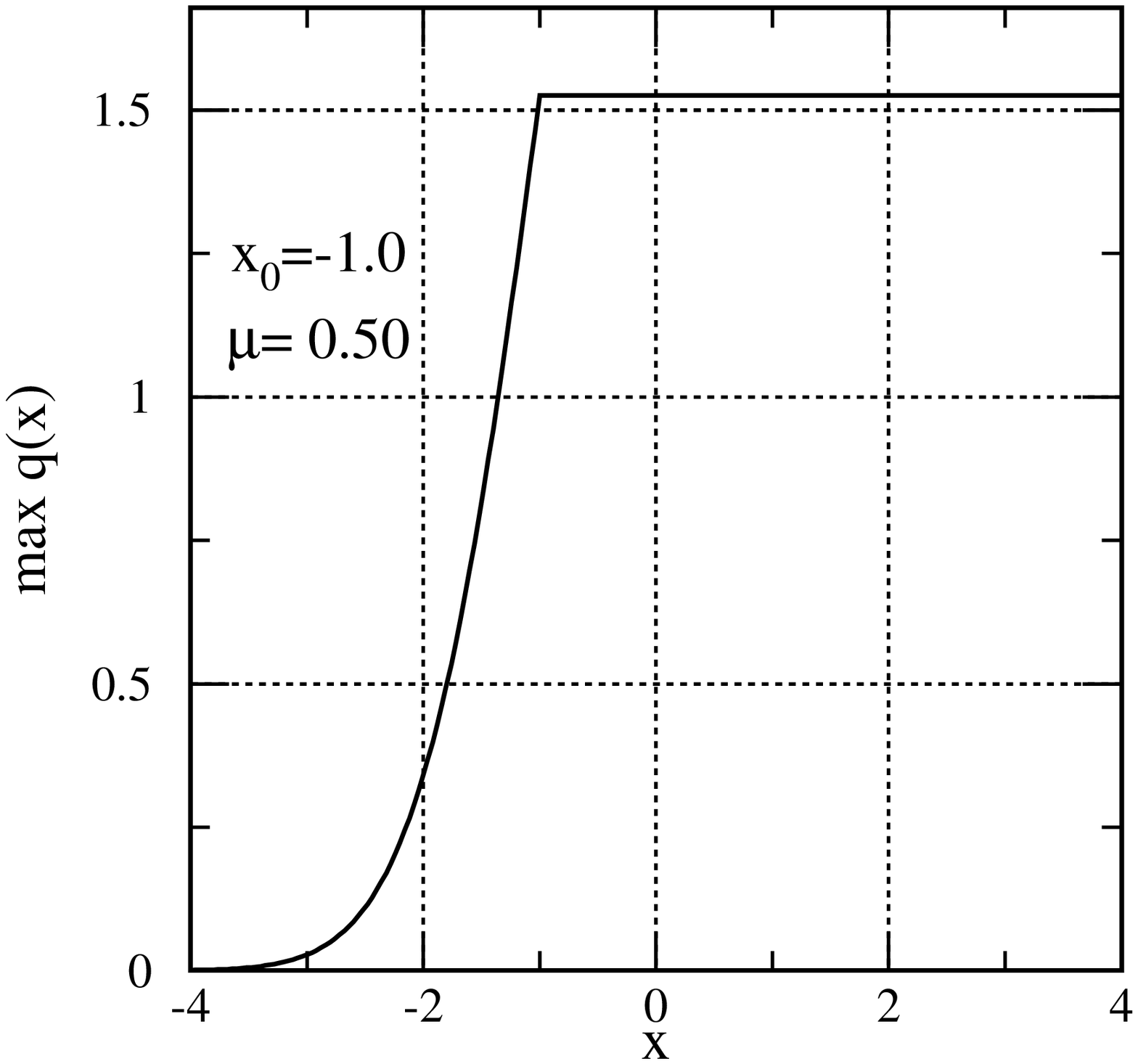} &
   \leavevmode \epsfxsize=5cm \epsfbox{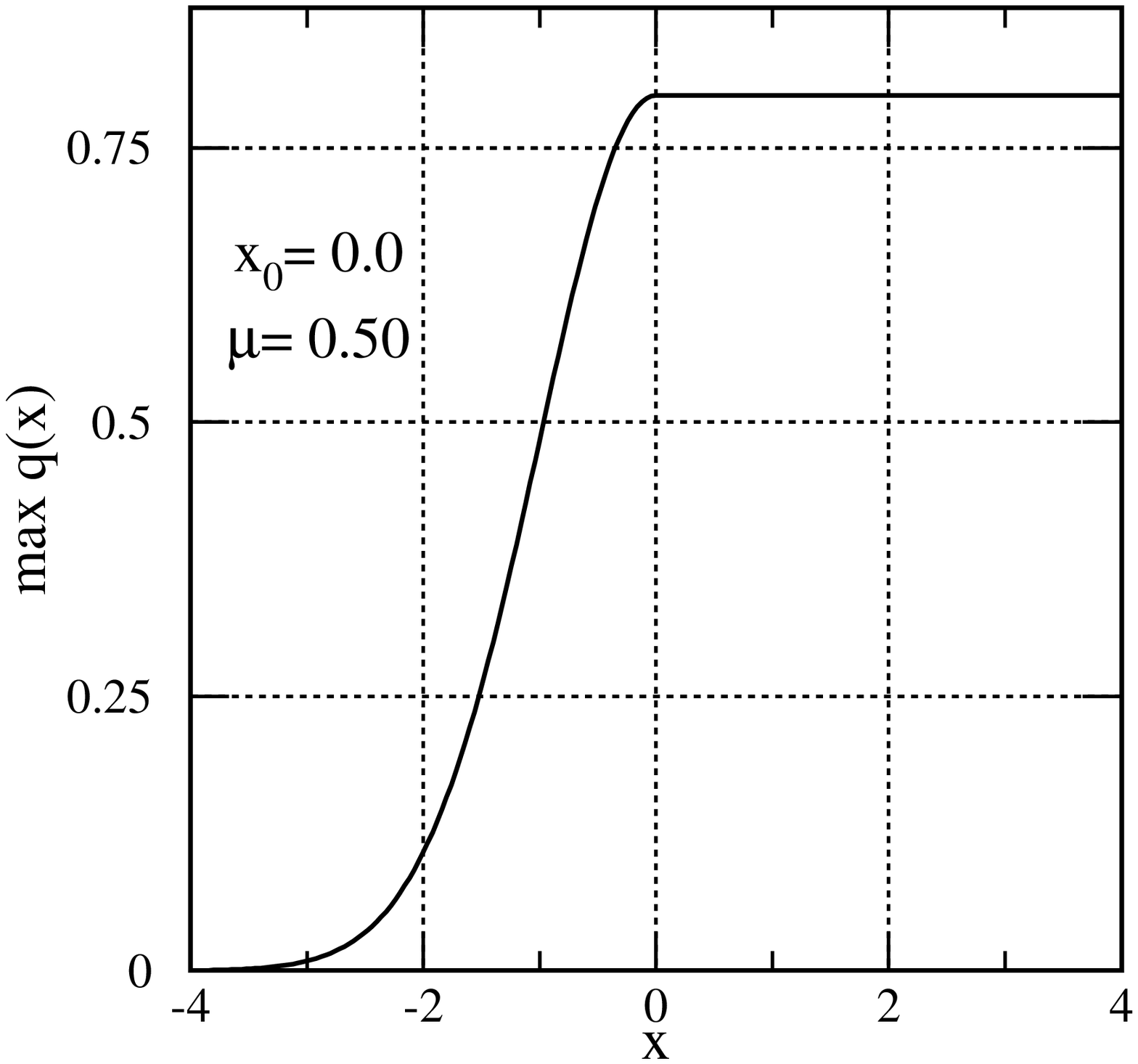} &
   \leavevmode \epsfxsize=5cm \epsfbox{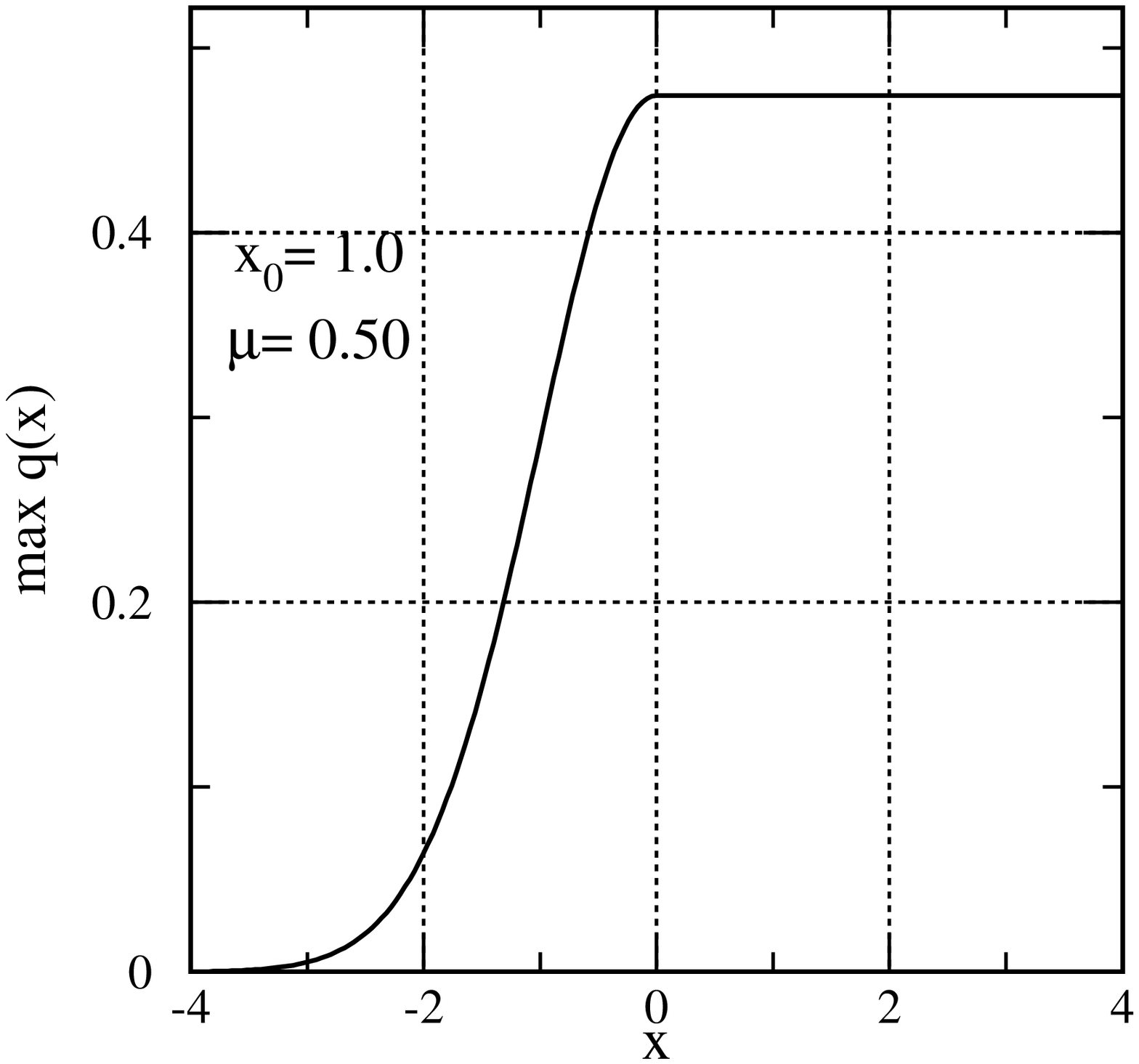} \cr
   \leavevmode \epsfxsize=5cm \epsfbox{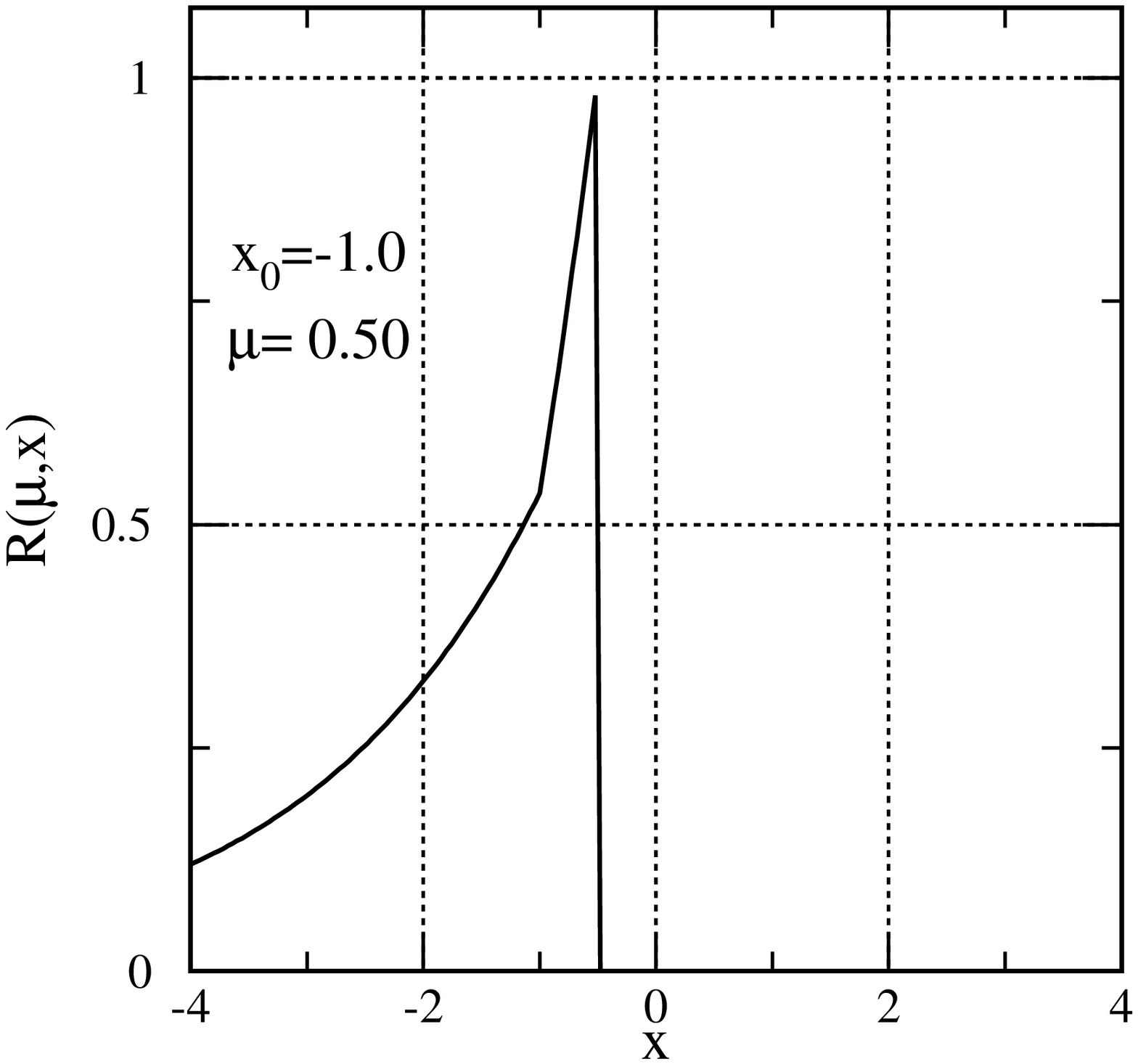} &
   \leavevmode \epsfxsize=5cm \epsfbox{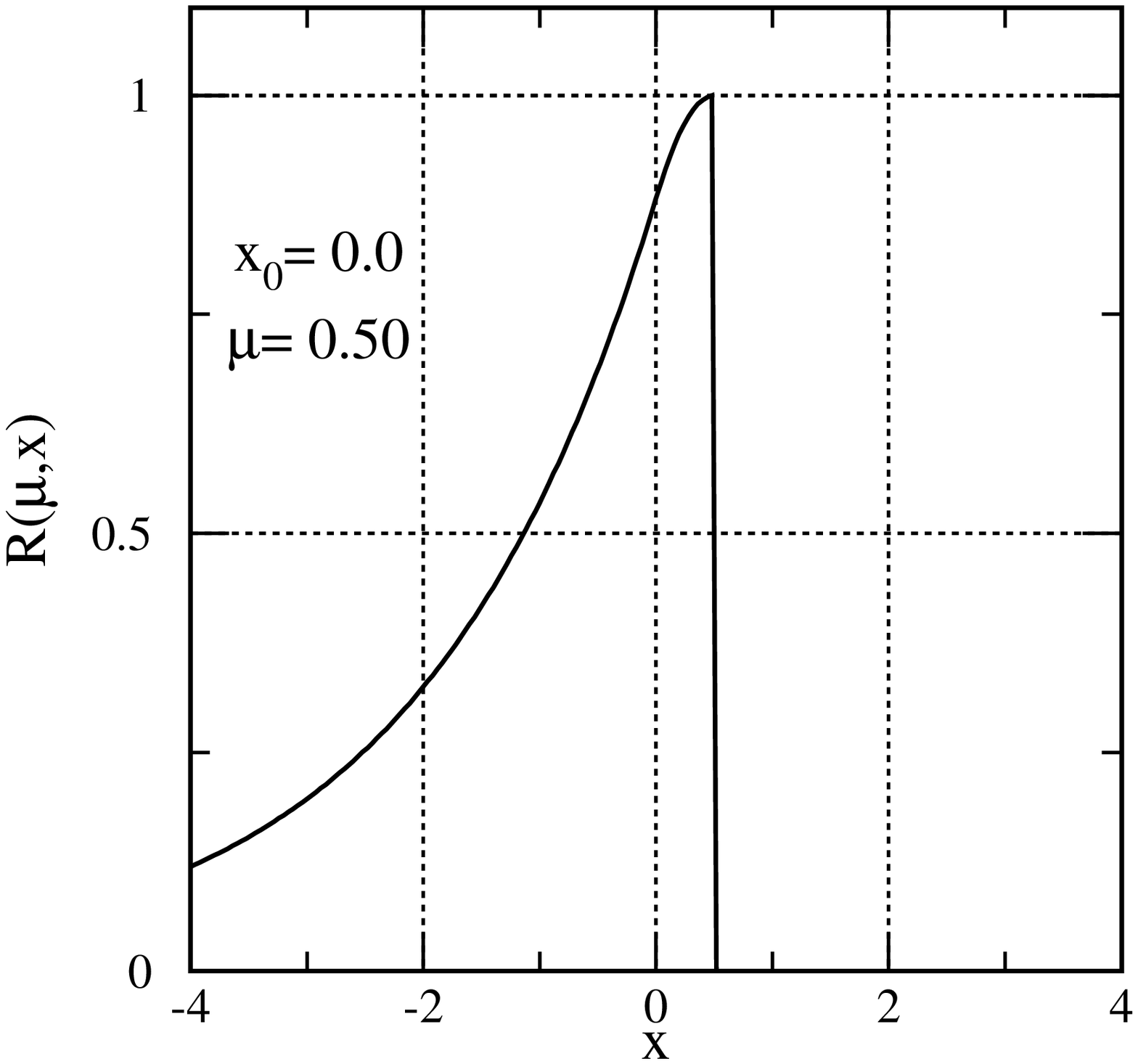} &
   \leavevmode \epsfxsize=5cm \epsfbox{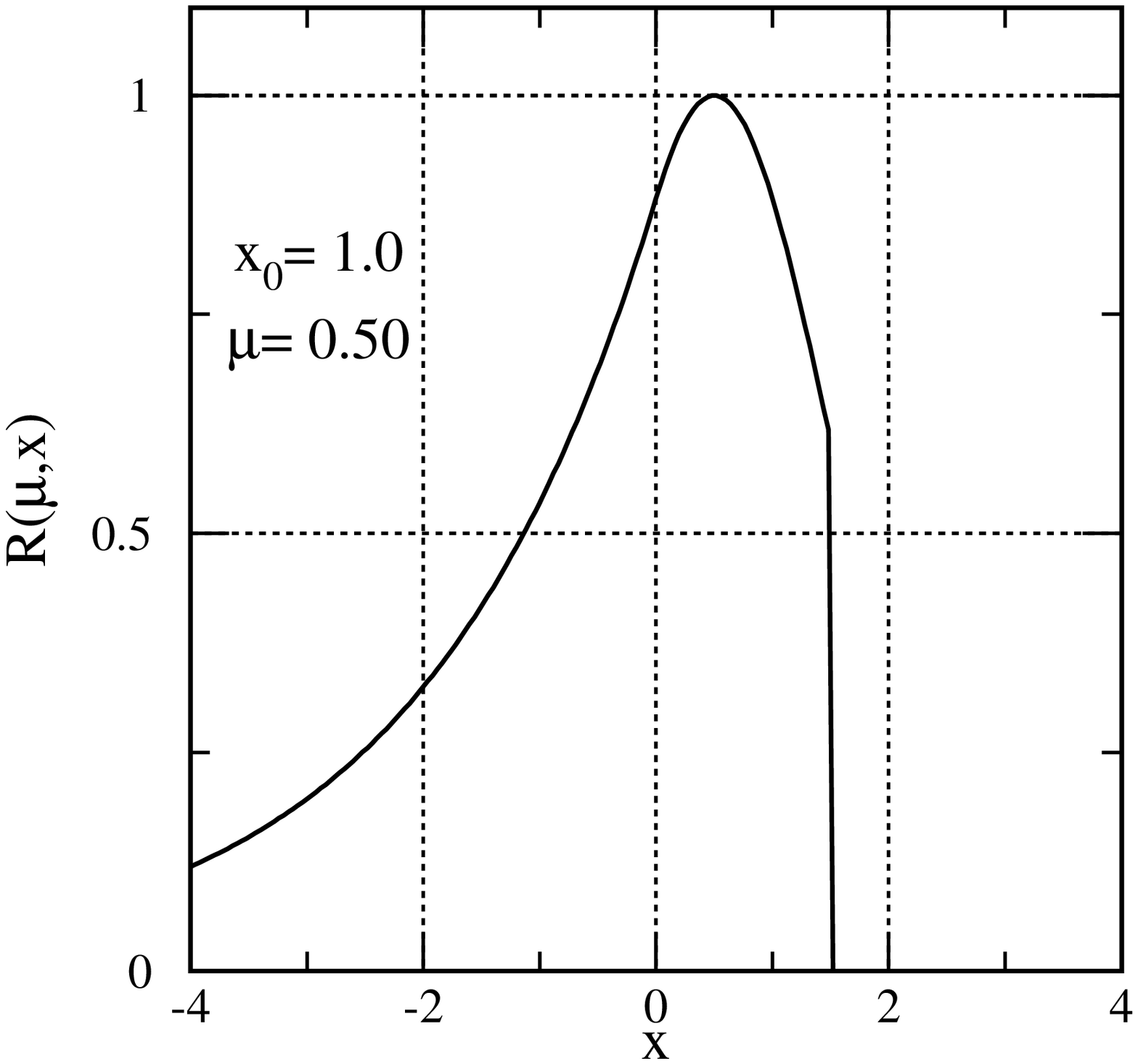} \cr
  \end{tabular}
\end{center}
\caption{Graphs of $ q^{x_0}_\mu(x)$ (top row), 
$\max_{\mu^\prime}\,q^{x_0}_{\mu^\prime}(x)$ (middle row), 
and $\Rnew^{x_0}(\mu,x)$ (bottom row), for $\mu=0.5$.
The columns are for $x_0$ = $-$1, 0, and 1.  Each graph in the bottom
row is the quotient of the two graphs above it.}
\label{fig-mu0.5}
\end{figure}

\begin{figure}
\begin{center}
  \begin{tabular}{ccc}
   \leavevmode \epsfxsize=5cm \epsfbox{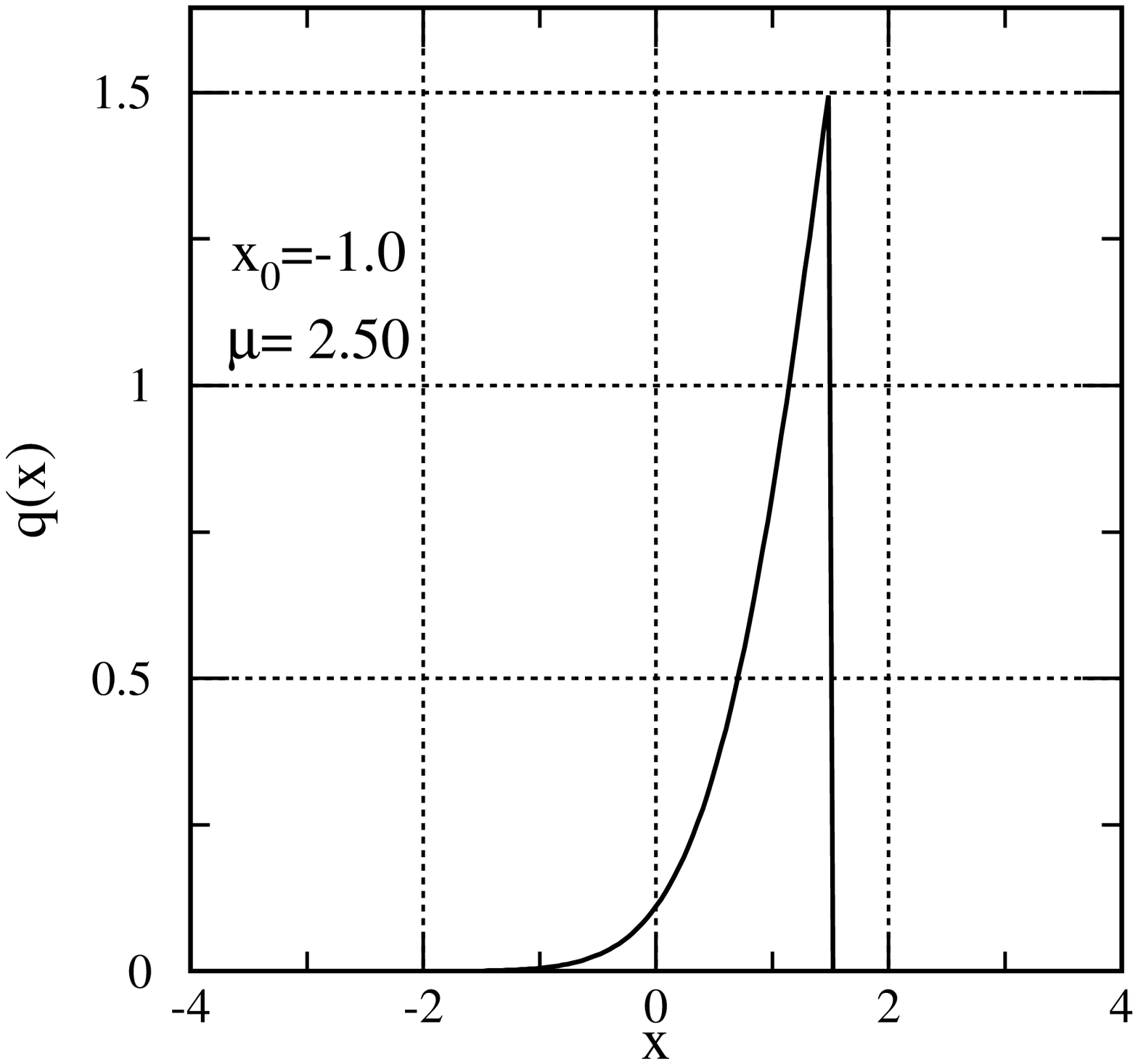} &
   \leavevmode \epsfxsize=5cm \epsfbox{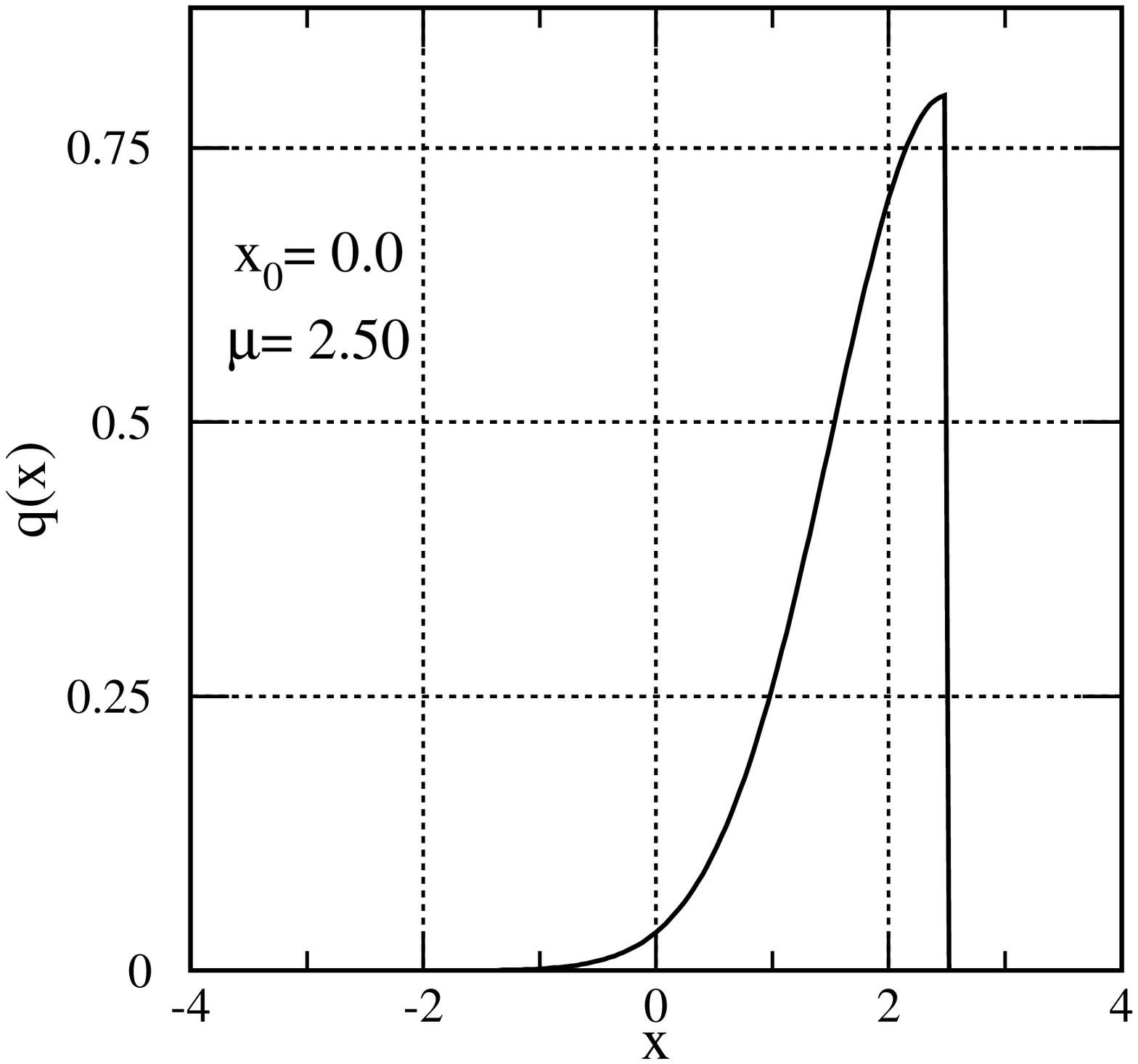} &
   \leavevmode \epsfxsize=5cm \epsfbox{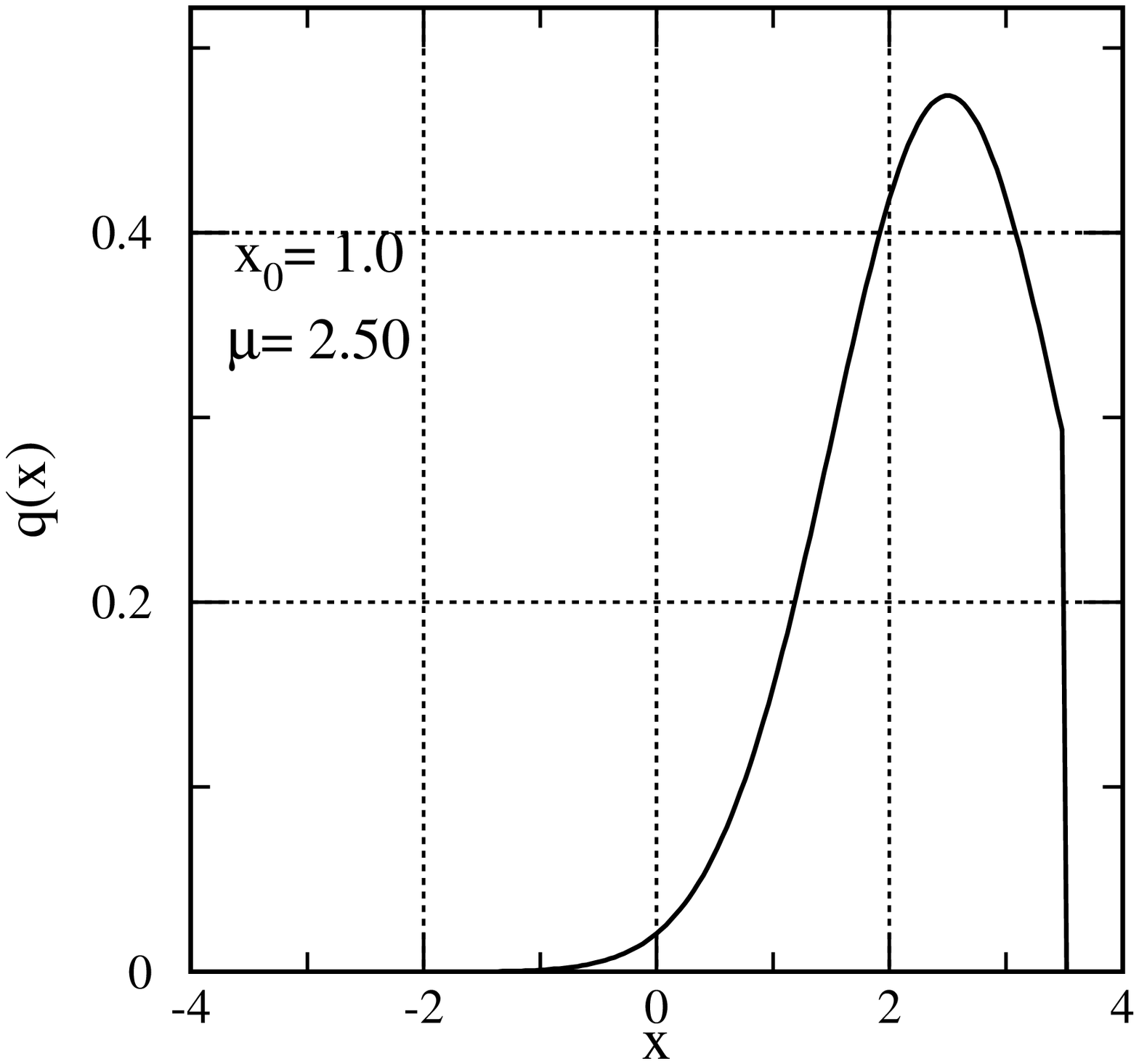} \cr
   \leavevmode \epsfxsize=5cm \epsfbox{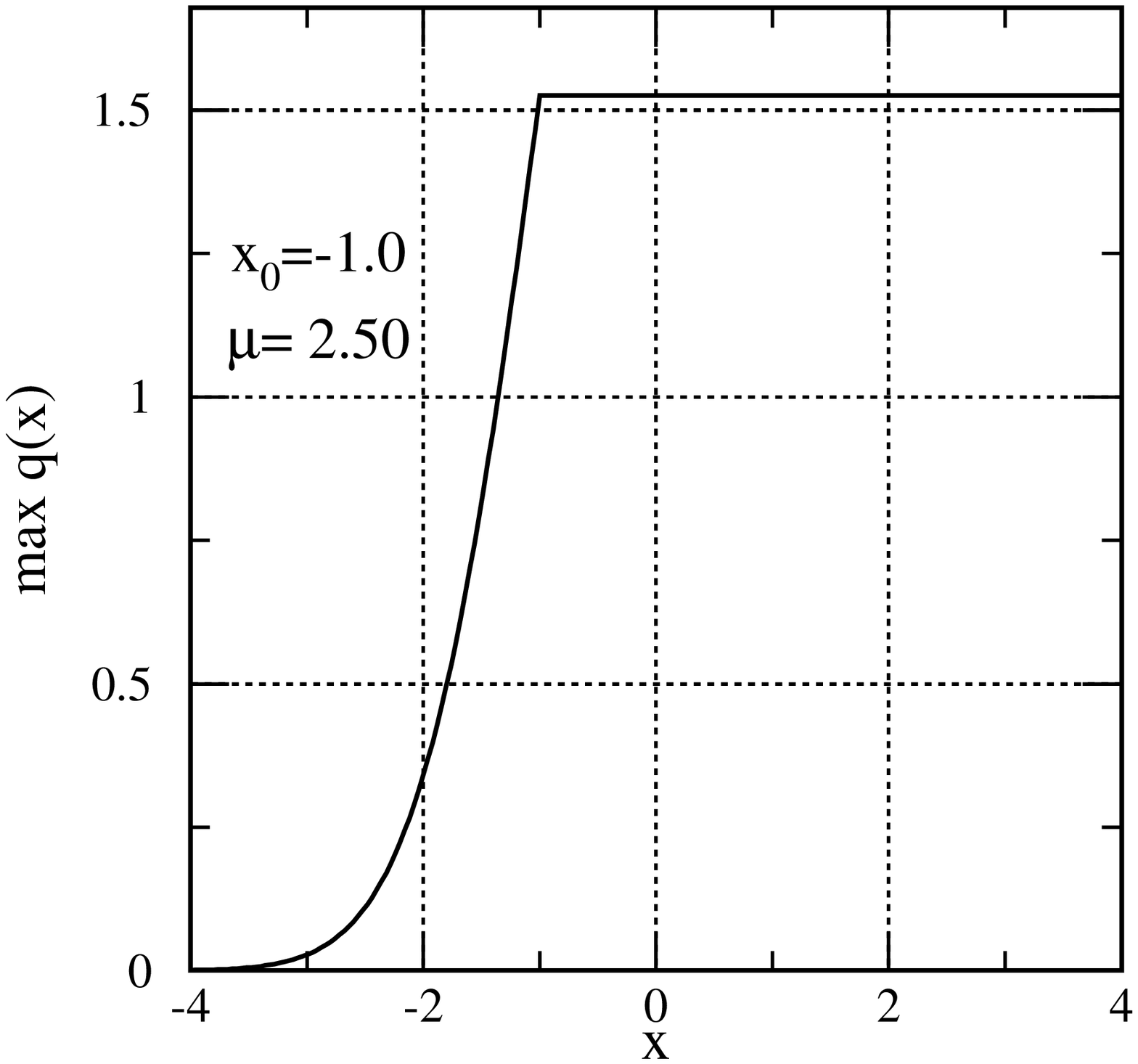} &
   \leavevmode \epsfxsize=5cm \epsfbox{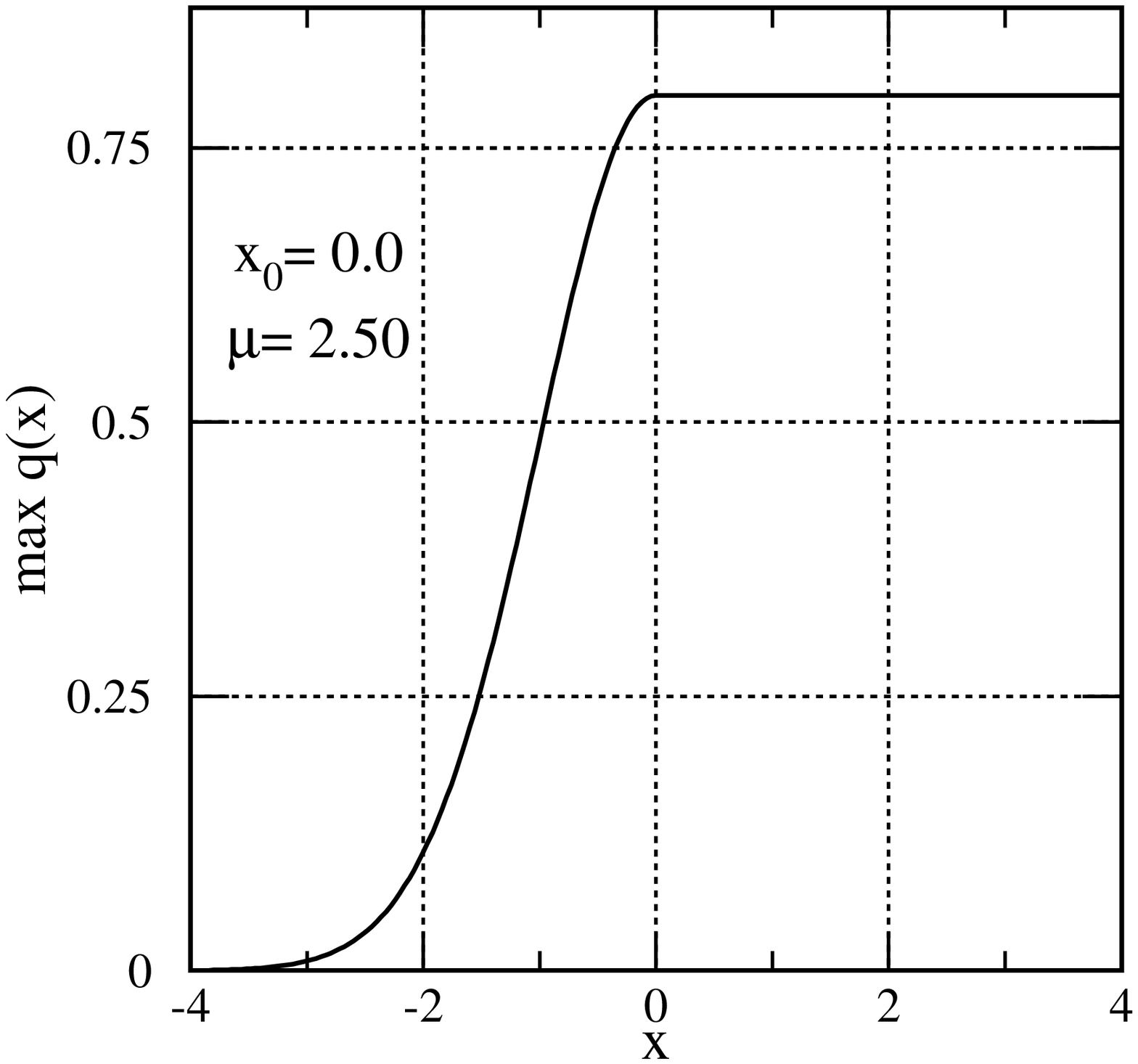} &
   \leavevmode \epsfxsize=5cm \epsfbox{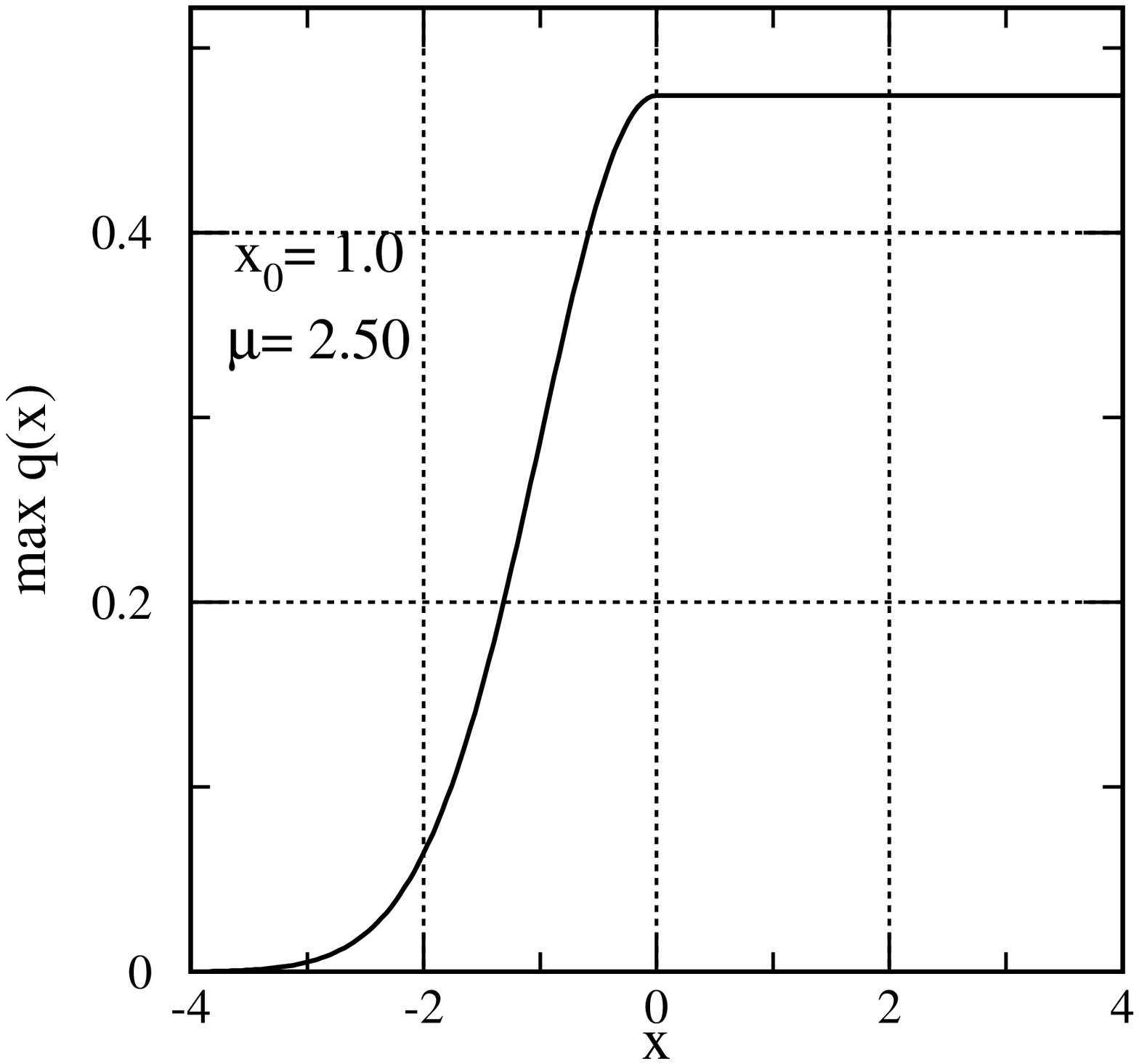} \cr
   \leavevmode \epsfxsize=5cm \epsfbox{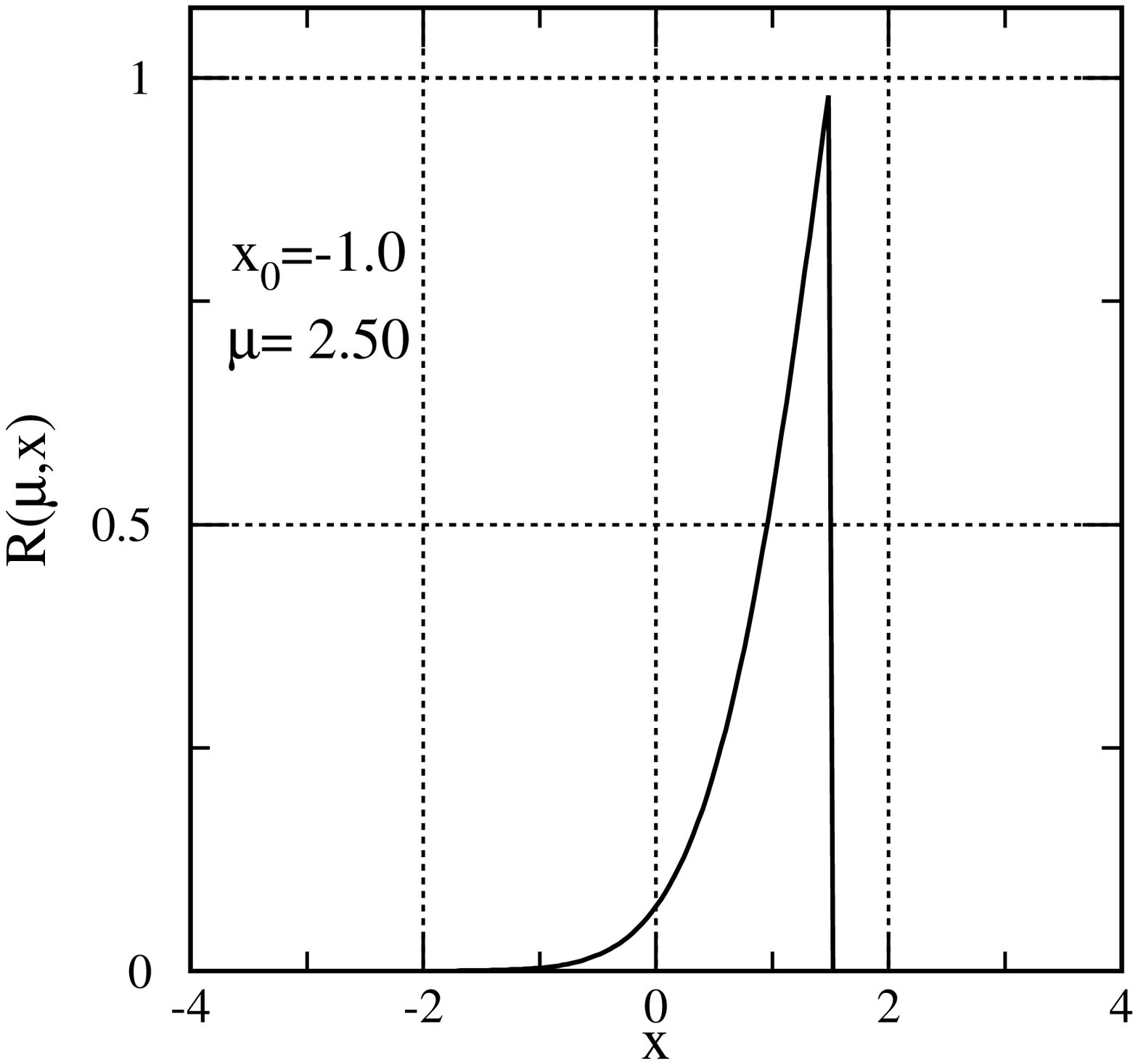} &
   \leavevmode \epsfxsize=5cm \epsfbox{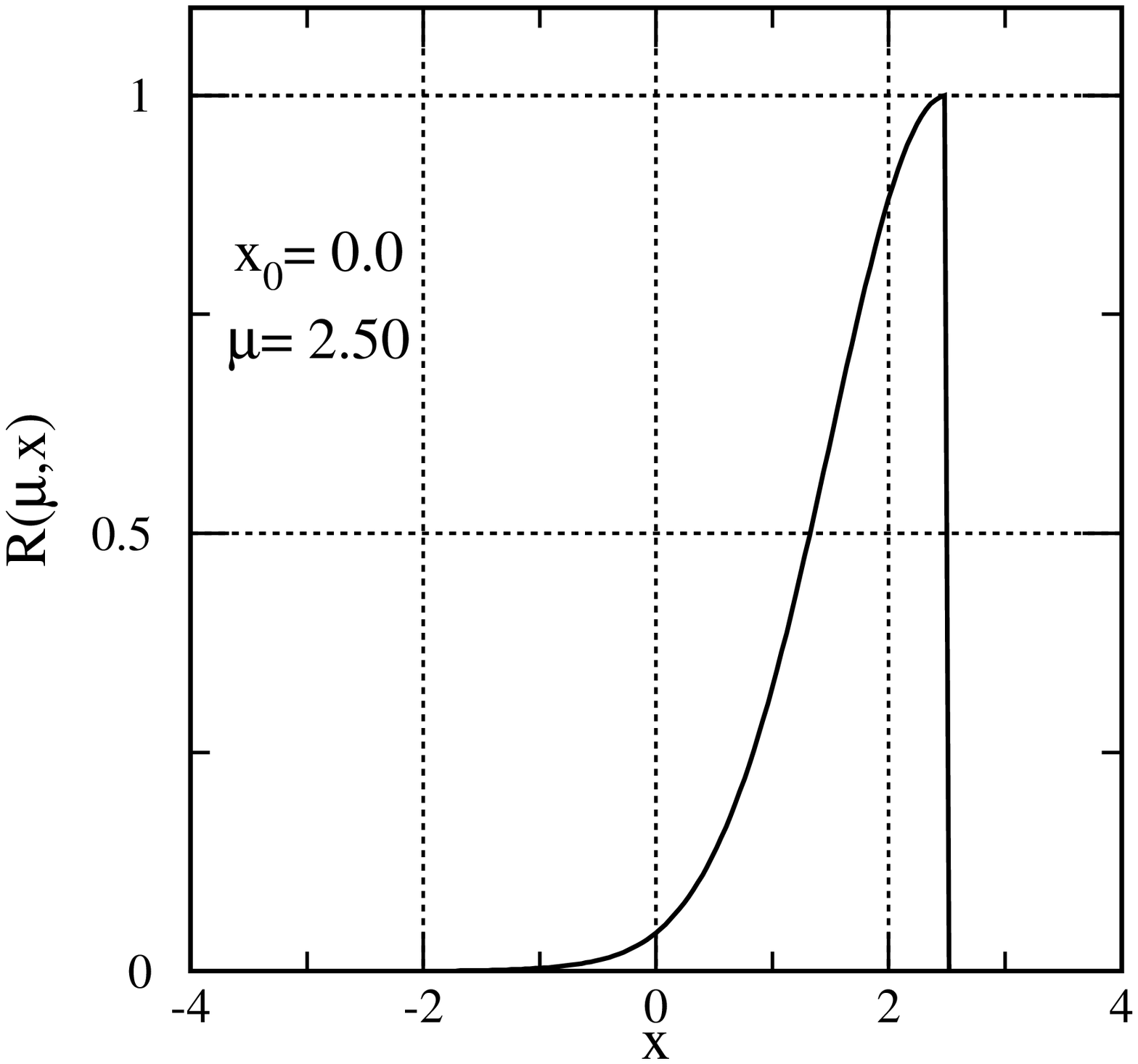} &
   \leavevmode \epsfxsize=5cm \epsfbox{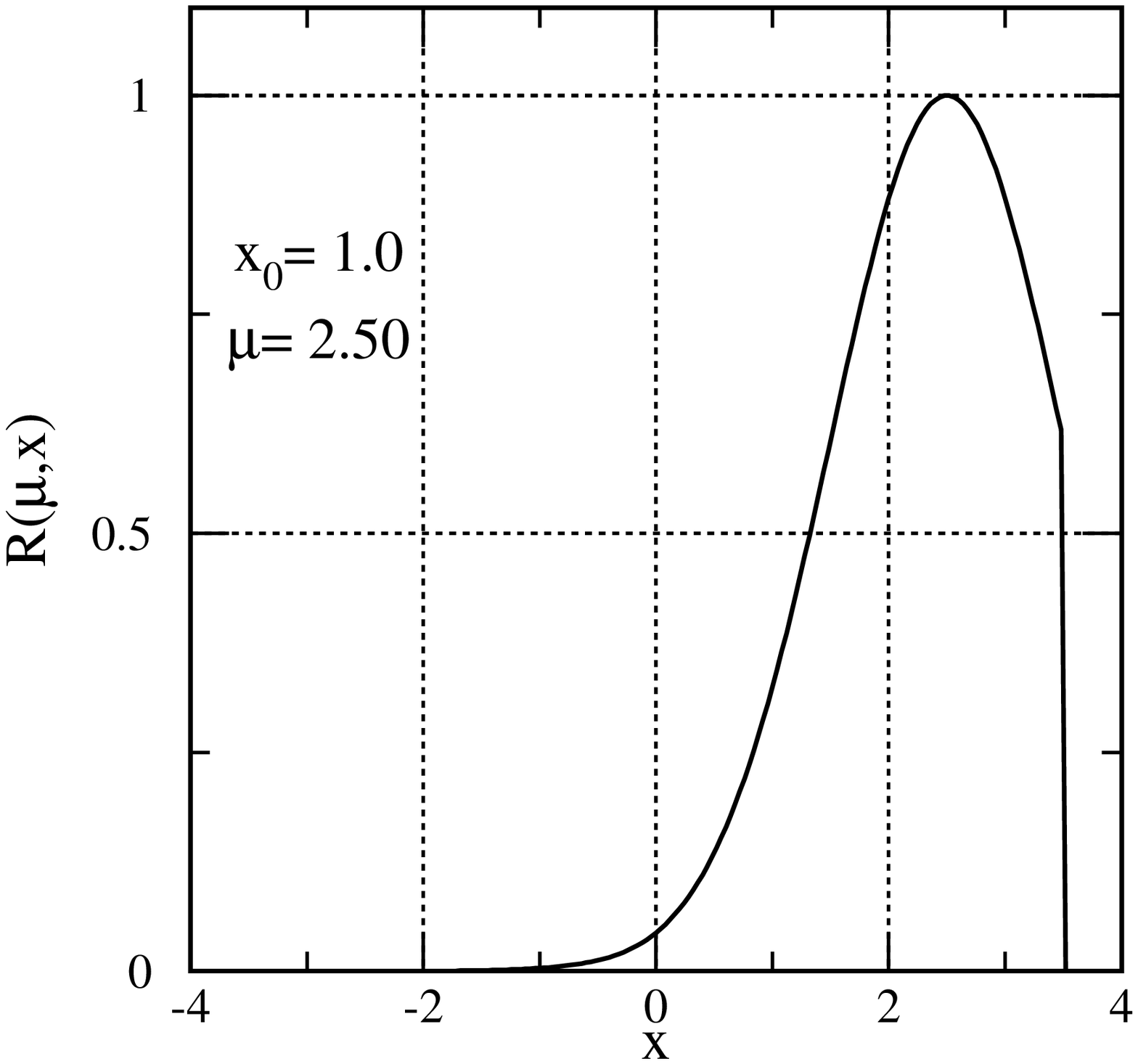} \cr
  \end{tabular}
\end{center}
\caption{Graphs of $ q^{x_0}_\mu(x)$ (top row), 
$\max_{\mu^\prime}\,q^{x_0}_{\mu^\prime}(x)$ (middle row), 
and $\Rnew^{x_0}(\mu,x)$ (bottom row), for $\mu=2.5$.
The columns are for $x_0$ = $-$1, 0. and 1.  Each graph in the bottom
row is the quotient of the two graphs above it.}
\label{fig-mu2.5}
\end{figure}

We let $P(x|\mu,\noise \le x_0)$ be the (normalized) conditional
probability for obtaining $x$, given that $\noise \le x_0$.  In
notation similar to that of RW, this can be denoted as
$q^{x_0}_\mu(x)$:
\begin{equation}
\label{eqn-gauss-cond}
q^{x_0}_\mu(x) \equiv P(x|\mu,\noise \le x_0)  
 = \left\{\begin{array}{ll}
          \frac{2}{\sqrt{2\pi}} \exp(-(x-\mu)^2/2)/
             (\erf(x_0/\sqrt{2})+1), 
              & \mbox{$x \le \mu + x_0$}\\
           0, & \mbox{$x  >  \mu + x_0$.}
          \end{array}
   \right.
\end{equation}
Given $x_0$, at each $x$ we find $\mubest$, that value of $\mu$ which
maximizes $P(x|\mu,\noise \le x_0)$:
\begin{equation}
\mubest = \left\{\begin{array}{ll}
                   x,       & \mbox{$x_0\ge0$ and $x\ge0$}\\
                   x - x_0, & \mbox{$x_0 < 0$ and $x\ge x_0$}\\
                   0,       & \mbox{otherwise}
                 \end{array}
          \right.
\end{equation}
In the notation of Ref.~\cite{RW}, $P(x|\mubest,\noise \le x_0)$ is
then
\begin{equation}
    \max_{\mu^\prime}\,q^{x_0}_{\mu^\prime}(x) =
          \frac{2}{\sqrt{2\pi} (\erf(x_0/\sqrt{2})+1)} \times
          \left\{\begin{array}{ll}
           1,               & \mbox{$x_0\ge0$ and $x\ge0$}\\
           \exp(-x_0^2)/2,  & \mbox{$x_0 < 0$ and $x\ge x_0$}\\
           \exp(-x^2)/2,    & \mbox{otherwise}
                 \end{array}
          \right.
 \end{equation}
Then the ratio $R$ of Eqn.~\ref{eqn-R-gauss} is replaced by
\begin{equation}
\label{eqn-R-tilde}
\Rnew^{x_0}(\mu,x) = {q^{x_0}_\mu(x) \over
    \max_{\mu^\prime}\,q^{x_0}_{\mu^\prime}(x)},
\end{equation}
which vanishes if $x > \mu + x_0$, and otherwise is given by
\begin{equation}
\Rnew^{x_0}(\mu,x) =
      \left\{\begin{array}{ll}
      \exp(-(x-\mu)^2/2),           & \mbox{$x_0\ge0$ and $x\ge0$}\\
      \exp((-(x-\mu)^2 + x_0^2)/2), & \mbox{$x_0 < 0$ and $x\ge x_0$}\\
      \exp(x\mu - \mu^2/2),         & \mbox{otherwise}
             \end{array}
      \right.
\end{equation}
Figures \ref{fig-mu0.01} through \ref{fig-mu2.5}
show graphs of $q^{x_0}_\mu(x)$, 
$\max_{\mu^\prime}\,q^{x_0}_{\mu^\prime}(x)$,
and $\Rnew^{x_0}(\mu,x)$, for three values of $\mu$, for each 
of three values of $x_0$.

We let $\Cnew_{x_0}(\mu)$ be the value of $c$ for which
\begin{equation}
\int_{x: \Rnew^{x_0}(\mu,x) < c} q^{x_0}_\mu(x) dx = \alpha.
\end{equation}
The modified confidence interval consists of those $\mu$ for which
\begin{equation}
 \Rnew^{x_0}(\mu,x_0) \ge  \Cnew_{x_0}(\mu).
\end{equation}

\begin{figure}
\begin{center}
  \begin{tabular}{cc}
   \leavevmode \epsfxsize=7cm \epsfbox{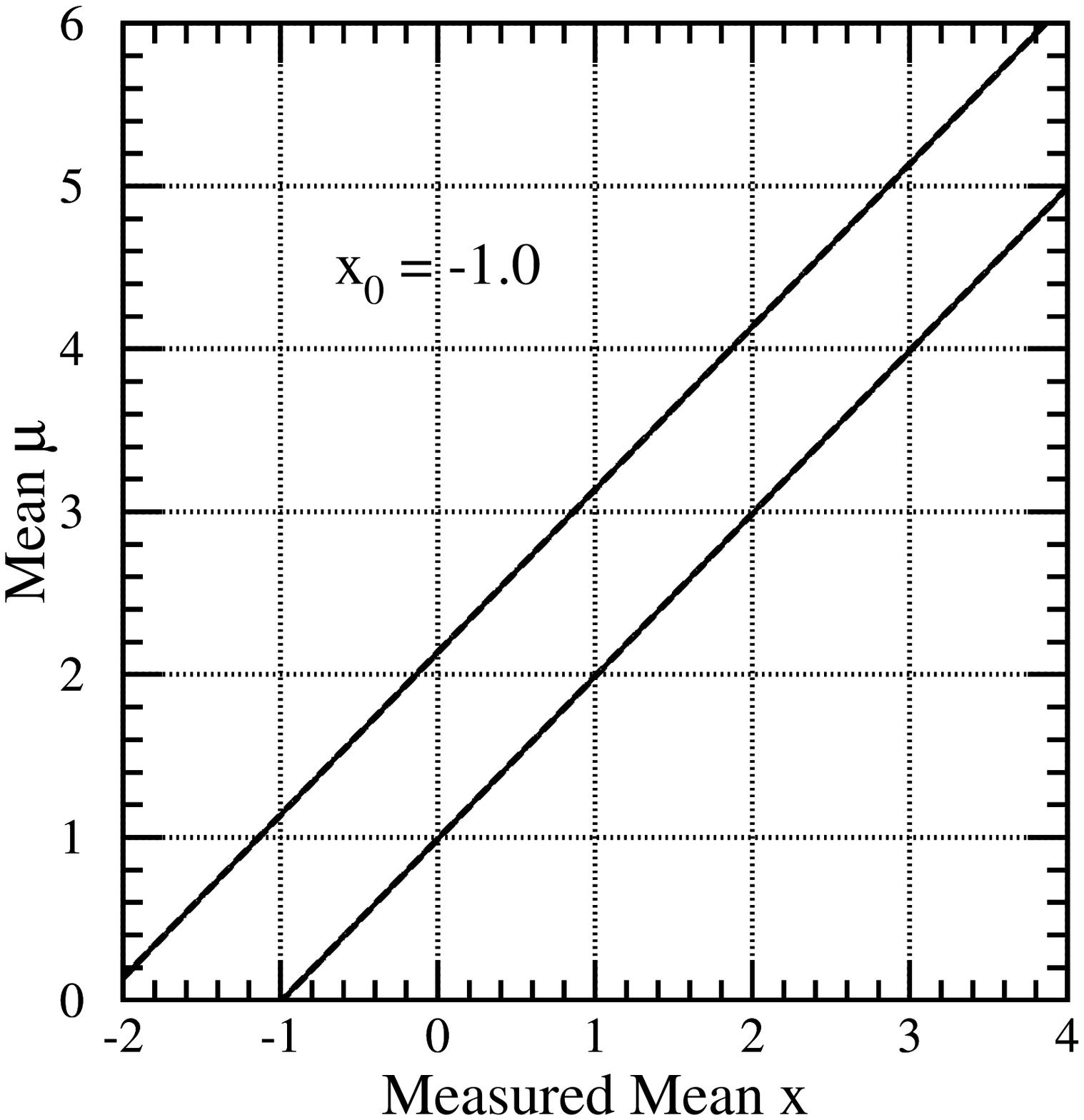} &
   \leavevmode \epsfxsize=7cm \epsfbox{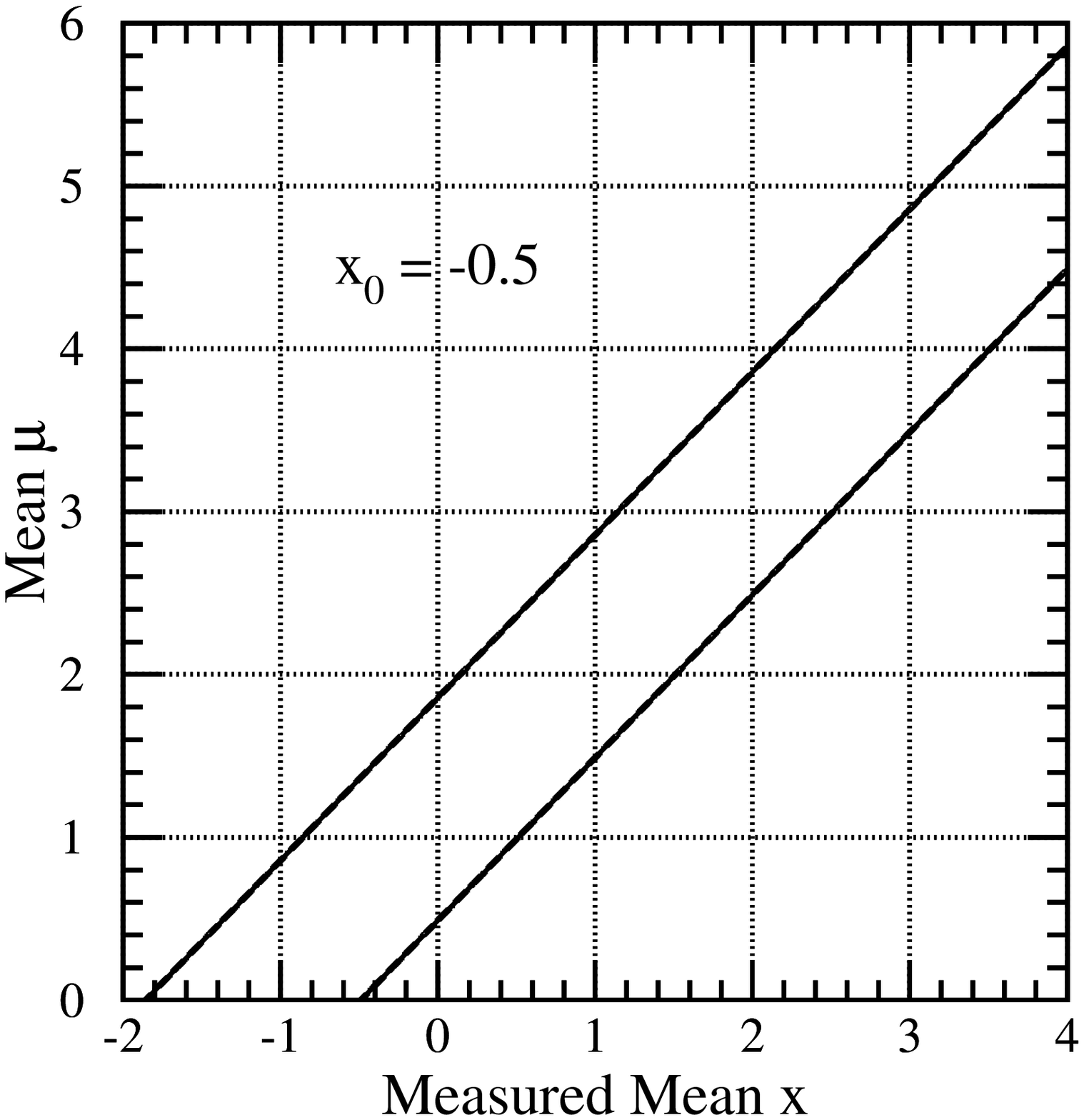} \cr
   \leavevmode \epsfxsize=7cm \epsfbox{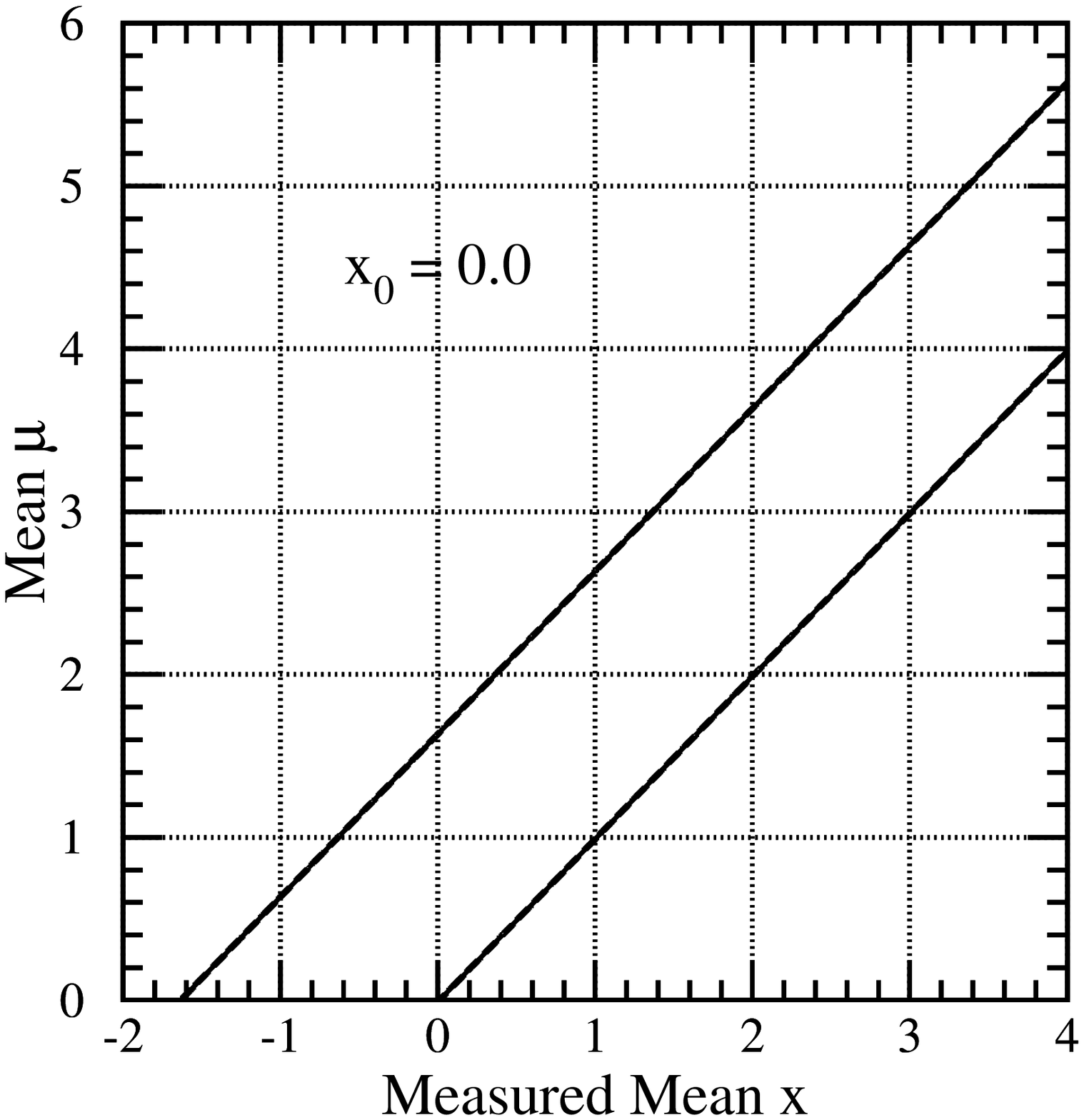} &
   \leavevmode \epsfxsize=7cm \epsfbox{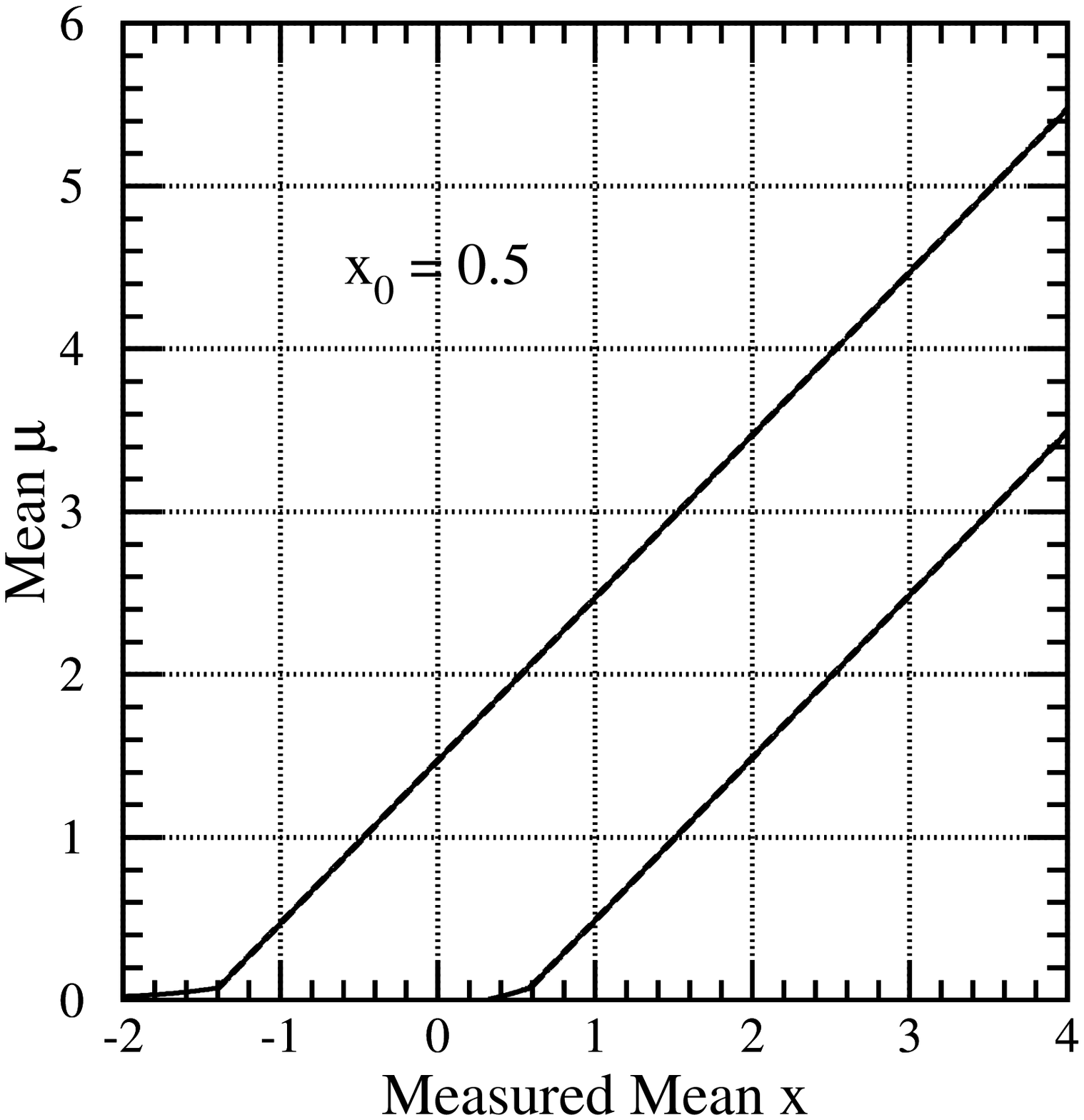} \cr
   \leavevmode \epsfxsize=7cm \epsfbox{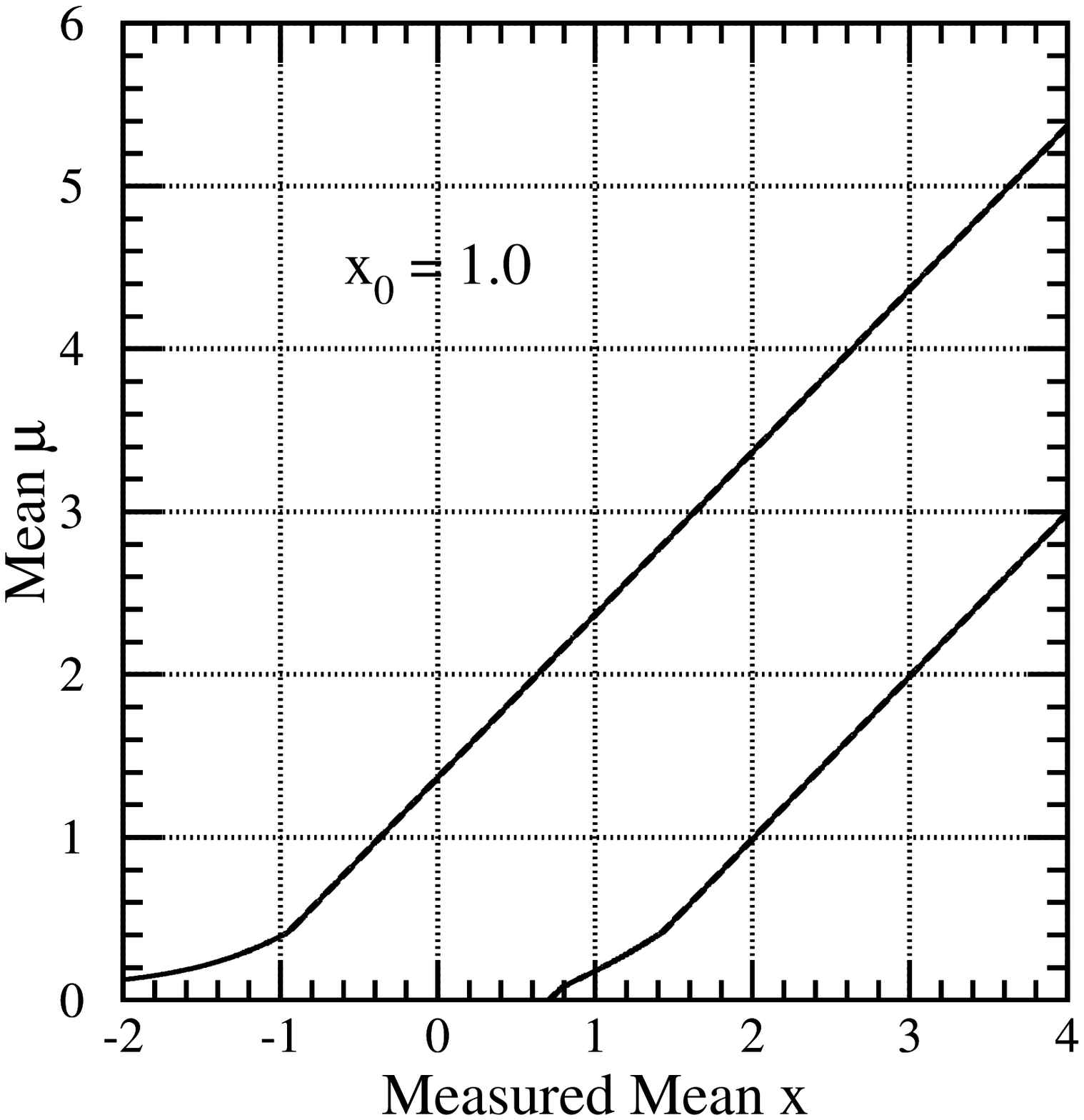} &
   \leavevmode \epsfxsize=7cm \epsfbox{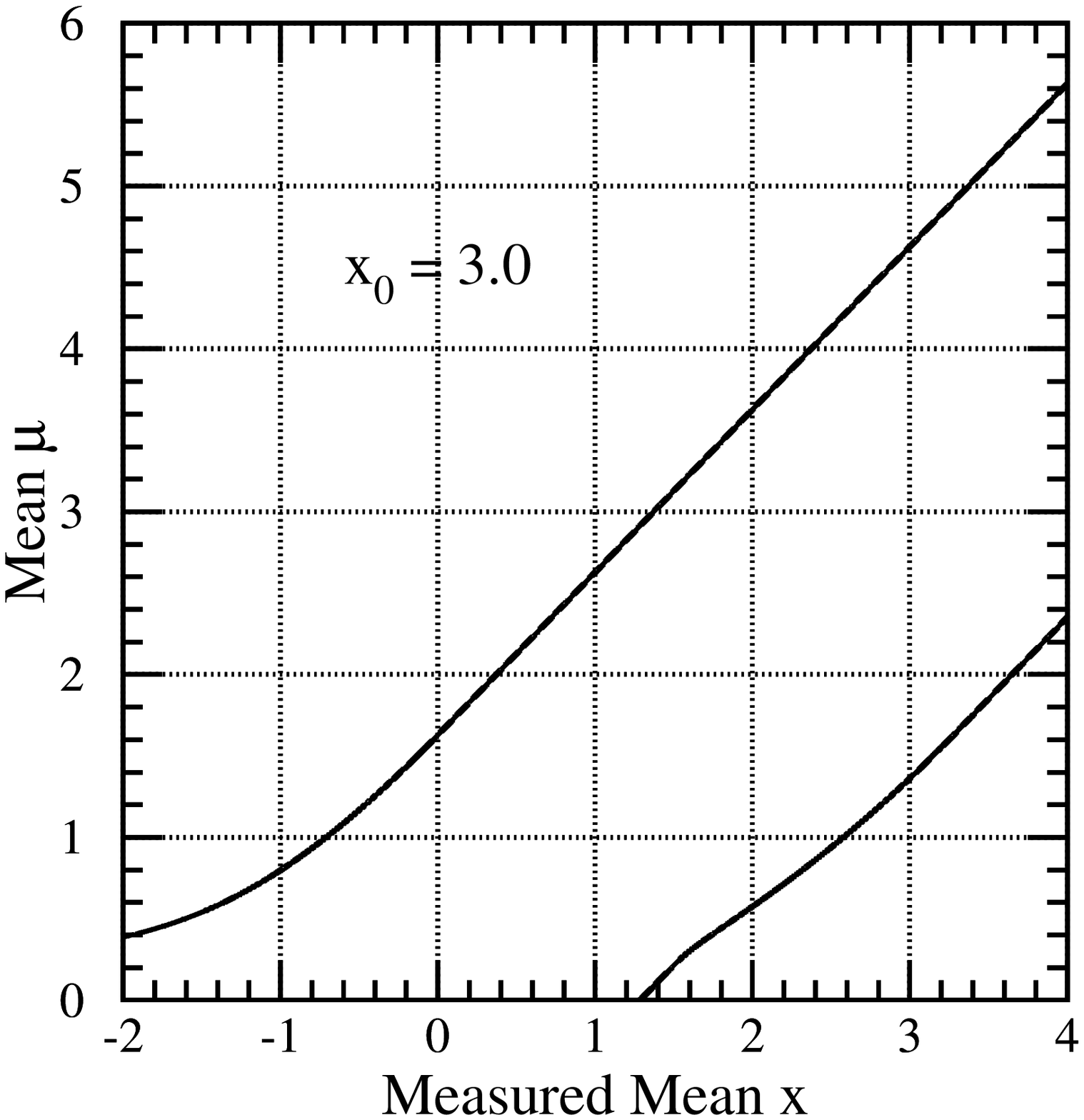} \cr
  \end{tabular}
\end{center}
\caption{Conditional confidence belts for the six sample values
of $x_0$ indicated.  Each plot is used only to find the
$[\mu_1,\mu_2]$ interval at $x$ equal to the $x_0$ used to construct
it; that interval is transferred to Fig.\protect\ref{fig-gauss-roe}.}
\label{fig-belts}
\end{figure}

Note that this entire construction depends on the value of $x_0$
obtained by the particular experiment.  An experiment obtaining a
different value of $x_0$ will have a different function in
Eqn.~\ref{eqn-gauss-cond}, and hence a different confidence belt
construction.  Figure \ref{fig-belts} shows examples of such
constructions for six values of $x_0$.  The vertical axis gives the
endpoints of the confidence intervals.  Each different confidence belt
construction is used only for an experiment obtaining the value $x_0$
which was used to construct the belt.  The interval $[\mu_1,\mu_2]$ at
$x=x_0$ is read off for that experiment; the rest of that plot is not
used.

\begin{figure}
\begin{center}
\leavevmode
\epsfxsize=15cm
\epsfbox{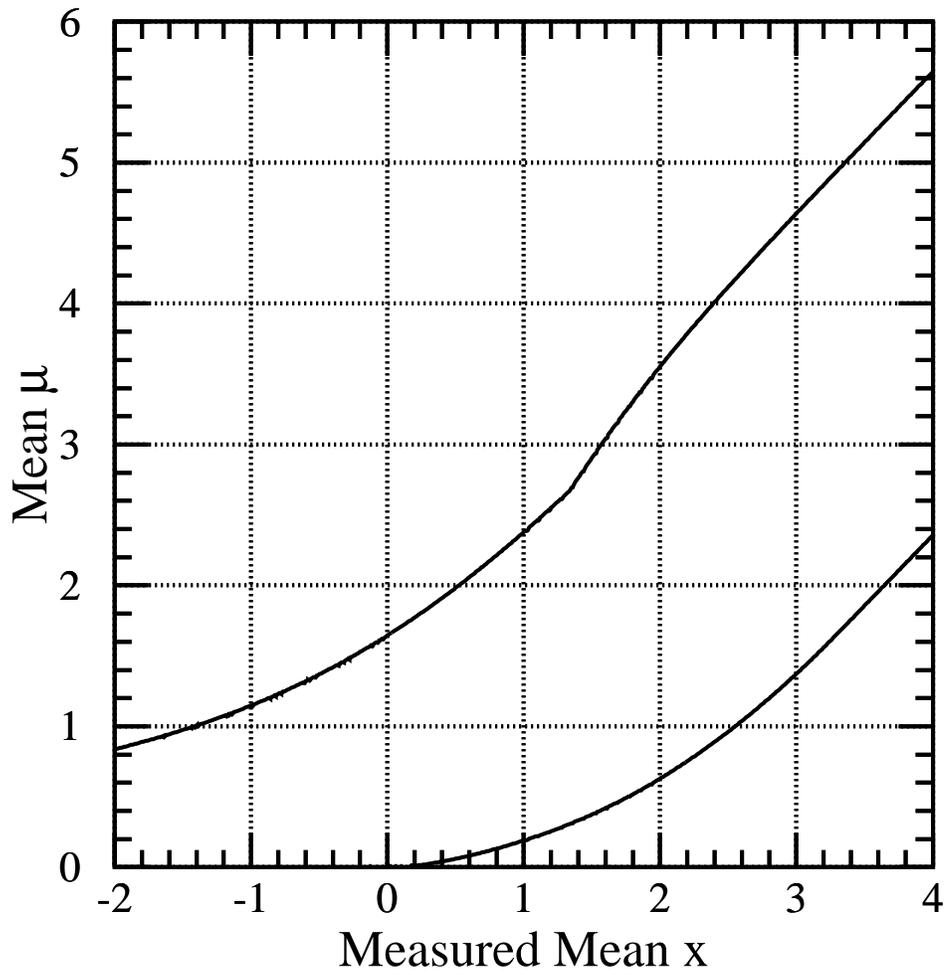}
\end{center}
\caption{Plot of RW-inspired 90\% conditional confidence intervals 
for mean of a Gaussian, constrained to be non-negative, described in
the text.  }
\label{fig-gauss-roe}
\end{figure}

Finally, we can form the graph shown in Fig.~\ref{fig-gauss-roe} by
taking the modified confidence interval for each $x_0$, and plotting
them all on one plot.  These are tabulated in
Table~\ref{tab-gauss-roe}, which includes for comparison the
unconditioned intervals from Table X of Ref.~\cite{FC}.

\begin{figure}
\begin{center}
\leavevmode
\epsfxsize=15cm
\epsfbox{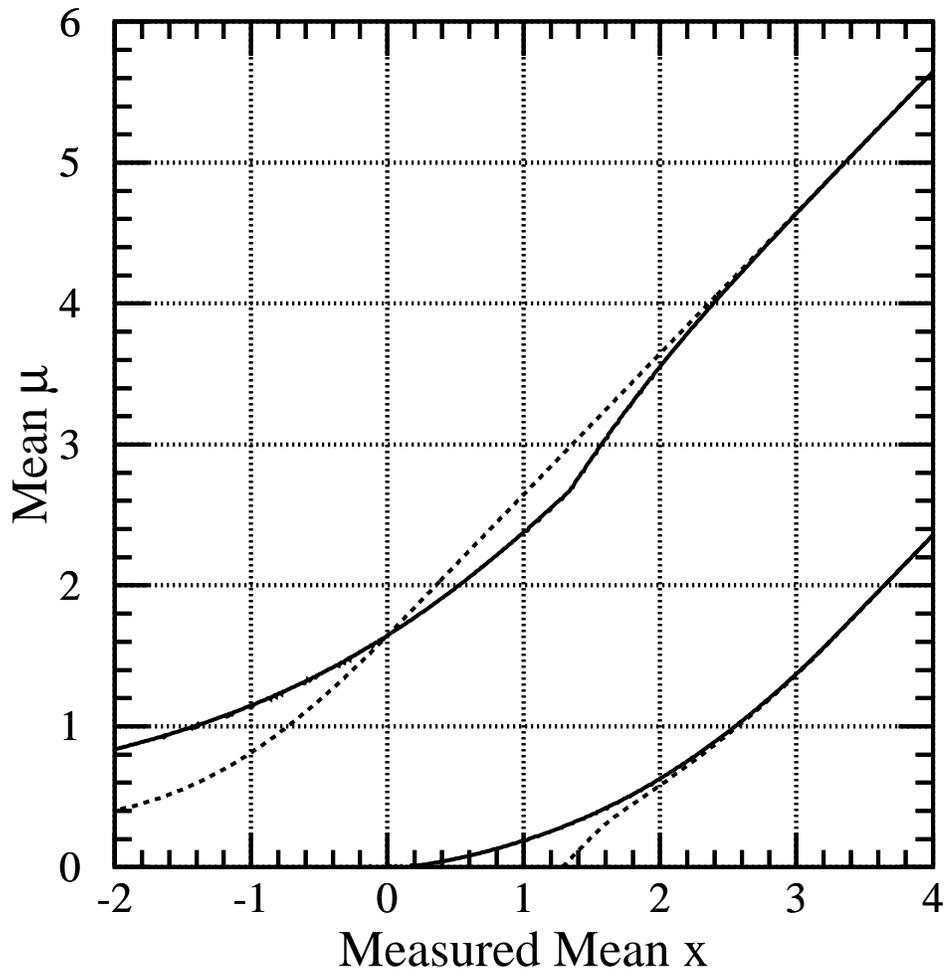}
\end{center}
\caption{Plot of RW-inspired 90\% conditional confidence intervals
(solid curves) ,
superimposed on the unconditioned intervals of 
Ref.~\protect\cite{FC} (dotted curves).  }
\label{fig-gauss-roe-fc}
\end{figure}

Fig.~\ref{fig-gauss-roe-fc} shows the modified intervals plotted
together with the unified intervals of Ref.~\cite{FC}.  The modified
upper curve is shifted upward for negative $x$, which results in a
less stringent upper limit when $\noise$ is known to be negative; this
feature is considered desirable by some.  The lower curve, however, is
also shifted upward: for all $x_0>0$, the interval is two-sided.  We
find this to be a highly undesirable side-effect.

It is interesting to consider what happens if one applies
Fig.~\ref{fig-gauss-roe} to an unconditioned ensemble.  The result can
be seen by drawing a horizontal line at any $\mu$ in
Fig.~\ref{fig-gauss-roe} and integrating $P(x|\mu)$
(Eqn.\ref{eqn-gauss}) along that line between the belts.  For small
$\mu$, there is significant undercoverage, while for $\mu$ near 1.0,
there is significant overcoverage.  The undercoverage was surprising,
since the conditioned intervals always cover within the relevant
subset of the ensemble.  However, conditioning on an inequality means
that these subsets are not disjoint.

The undesirable raising of the lower curve is present in the Poisson
case, as can be seen in Figure 1 of Ref.~\cite{RW}.  However, there
the discreteness of the Poisson problem apparently prevents the curve
from being shifted so dramatically, and the two-sided intervals do not
extend to such low values of the measured $n$.

\section{Conclusion}
\label{sec-conclude}
In this paper, we apply conditioning in the style Roe and Woodroofe to
the Gaussian-with-boundary problem.  We find that the transition from
one-sided intervals to two-sided intervals undesirably moves to the
origin.  This reflects a general feature of confidence interval
construction: when moving one of the two curves, the other curve moves
also.  In the Poisson-with-background problem, the undesirable
movement was not large, but in the Gaussian-with-boundary problem, the
effect is quite substantial.

\acknowledgments
I thank Gary Feldman, Byron Roe, and Michael Woodroofe for comments on
the paper.  This work was supported by the U.S. Department of Energy.

\clearpage

\begin{table}
\caption{90\% C.L. confidence intervals 
for the mean $\mu$ of a Gaussian,
constrained to be non-negative, as a function of the measured mean
$x_0$, for the RW conditioning method, and for the
unified approach of Feldman and Cousins.  
All numbers are in units of $\sigma$. 
The conditioned numbers may be inaccurate at the level
of $\pm 0.01$ due to the computational grid used.}
\label{tab-gauss-roe}
\bigskip
\begin{tabular}{dcc} \hline
 $x_0$ & conditioned  & unconditioned    \\ \hline
  -3.0 &   ( 0.00, 0.63) &    0.00, 0.26 \\
  -2.9 &   ( 0.00, 0.66) &    0.00, 0.27 \\
  -2.8 &   ( 0.00, 0.68) &    0.00, 0.28 \\
  -2.7 &   ( 0.00, 0.68) &    0.00, 0.29 \\
  -2.6 &   ( 0.00, 0.70) &    0.00, 0.30 \\
  -2.5 &   ( 0.00, 0.73) &    0.00, 0.32 \\
  -2.4 &   ( 0.00, 0.75) &    0.00, 0.33 \\
  -2.3 &   ( 0.00, 0.77) &    0.00, 0.34 \\
  -2.2 &   ( 0.00, 0.78) &    0.00, 0.36 \\
  -2.1 &   ( 0.00, 0.80) &    0.00, 0.38 \\
  -2.0 &   ( 0.00, 0.84) &    0.00, 0.40 \\
  -1.9 &   ( 0.00, 0.86) &    0.00, 0.43 \\
  -1.8 &   ( 0.00, 0.89) &    0.00, 0.45 \\
  -1.7 &   ( 0.00, 0.92) &    0.00, 0.48 \\
  -1.6 &   ( 0.00, 0.94) &    0.00, 0.52 \\
  -1.5 &   ( 0.00, 0.97) &    0.00, 0.56 \\
  -1.4 &   ( 0.00, 1.01) &    0.00, 0.60 \\
  -1.3 &   ( 0.00, 1.04) &    0.00, 0.64 \\
  -1.2 &   ( 0.00, 1.07) &    0.00, 0.70 \\
  -1.1 &   ( 0.00, 1.11) &    0.00, 0.75 \\
  -1.0 &   ( 0.00, 1.15) &    0.00, 0.81 \\
  -0.9 &   ( 0.00, 1.19) &    0.00, 0.88 \\
  -0.8 &   ( 0.00, 1.23) &    0.00, 0.95 \\
  -0.7 &   ( 0.00, 1.27) &    0.00, 1.02 \\
  -0.6 &   ( 0.00, 1.32) &    0.00, 1.10 \\
  -0.5 &   ( 0.00, 1.37) &    0.00, 1.18 \\
  -0.4 &   ( 0.00, 1.42) &    0.00, 1.27 \\
  -0.3 &   ( 0.00, 1.47) &    0.00, 1.36 \\
  -0.2 &   ( 0.00, 1.53) &    0.00, 1.45 \\
  -0.1 &   ( 0.00, 1.58) &    0.00, 1.55 \\
   0.0 &   ( 0.00, 1.65) &    0.00, 1.64 \\
   0.1 &   ( 0.00, 1.71) &    0.00, 1.74 \\
   0.2 &   ( 0.01, 1.77) &    0.00, 1.84 \\
   0.3 &   ( 0.02, 1.84) &    0.00, 1.94 \\
   0.4 &   ( 0.04, 1.91) &    0.00, 2.04 \\
   0.5 &   ( 0.06, 1.98) &    0.00, 2.14 \\
   0.6 &   ( 0.08, 2.06) &    0.00, 2.24 \\
   0.7 &   ( 0.11, 2.13) &    0.00, 2.34 \\
   0.8 &   ( 0.13, 2.21) &    0.00, 2.44 \\
   0.9 &   ( 0.16, 2.29) &    0.00, 2.54 \\
   1.0 &   ( 0.19, 2.38) &    0.00, 2.64 \\
   1.1 &   ( 0.22, 2.46) &    0.00, 2.74 \\
   1.2 &   ( 0.26, 2.55) &    0.00, 2.84 \\
   1.3 &   ( 0.29, 2.64) &    0.02, 2.94 \\
   1.4 &   ( 0.33, 2.76) &    0.12, 3.04 \\
   1.5 &   ( 0.38, 2.90) &    0.22, 3.14 \\
   1.6 &   ( 0.42, 3.04) &    0.31, 3.24 \\
   1.7 &   ( 0.47, 3.18) &    0.38, 3.34 \\
   1.8 &   ( 0.52, 3.30) &    0.45, 3.44 \\
   1.9 &   ( 0.57, 3.43) &    0.51, 3.54 \\
   2.0 &   ( 0.63, 3.55) &    0.58, 3.64 \\
   2.1 &   ( 0.69, 3.67) &    0.65, 3.74 \\
   2.2 &   ( 0.75, 3.79) &    0.72, 3.84 \\
   2.3 &   ( 0.82, 3.90) &    0.79, 3.94 \\
   2.4 &   ( 0.89, 4.01) &    0.87, 4.04 \\
   2.5 &   ( 0.96, 4.12) &    0.95, 4.14 \\
   2.6 &   ( 1.04, 4.22) &    1.02, 4.24 \\
   2.7 &   ( 1.12, 4.33) &    1.11, 4.34 \\
   2.8 &   ( 1.20, 4.43) &    1.19, 4.44 \\
   2.9 &   ( 1.28, 4.54) &    1.28, 4.54 \\
   3.0 &   ( 1.37, 4.64) &    1.37, 4.64 \\
   3.1 &   ( 1.47, 4.74) &    1.46, 4.74 \\
\end{tabular}
\end{table}


\begin{references}
\bibitem[\dagger]{bylineRC} cousins@physics.ucla.edu


\bibitem{RW}
B.P. Roe and M.B. Woodroofe, Phys. Rev. {\bf D60} 053009 (1999).

\bibitem{FC}
G.J. Feldman and R.D. Cousins, Phys. Rev. {\bf D57} 3873 (1998).

\end{references}
\end{document}